\documentclass[11pt,draftcls,onecolumn]{IEEEtran}
\usepackage{amsfonts}
\usepackage{amssymb}
\usepackage{mathrsfs}
\usepackage{amsmath}
\usepackage{cite}
\usepackage{mathrsfs}
\usepackage[ruled,commentsnumbered, vlined]{algorithm2e}
\usepackage{tikz}
\usetikzlibrary{arrows,automata,decorations.pathmorphing,decorations.footprints,fadings,calc,trees,mindmap,shadows,decorations.text,patterns,positioning,shapes,matrix,fit}
\usepackage[latin1]{inputenc}
\usepackage{verbatim}
\usepackage{graphicx}
\usepackage{subfigure}												%
\usepackage{ifpdf}															%
\ifpdf																		%
	\usepackage{hyperref}
\else																		%
\fi	
\usetikzlibrary{arrows,decorations.pathmorphing,decorations.footprints,fadings,calc,trees,mindmap,shadows,decorations.text,patterns,positioning,shapes,matrix,fit}

\makeatletter

\newcommand{\Rmnum}[1]{\expandafter\@slowromancap\romannumeral #1@}
\makeatother

\newtheorem{thm}{\bf Theorem}
\newtheorem{lemma}[thm]{\bf Lemma}
\newtheorem{eg}{\bf Example}
\newtheorem{prop}[thm]{\bf Proposition}
\newtheorem{cor}[thm]{\bf Corollary}

\newtheorem{defn}{\bf Definition}
\newtheorem{rem}[thm]{\bf Remark}

\newtheorem{claim}{\bf Claim}

\newcommand{\mA}{\mathscr{A}}
\newcommand{\mB}{\mathscr{B}}

\newcommand{\mO}{\mathcal{O}}
\newcommand{\mK}{\mathcal{K}}
\newcommand{\mM}{\mathcal{M}}
\newcommand{\mF}{\mathcal{F}}

\newcommand{\tail}{{\mathrm{tail}}}
\newcommand{\head}{{\mathrm{head}}}
\newcommand{\Out}{{\mathrm{Out}}}
\newcommand{\In}{{\mathrm{In}}}

\newcommand{\CUT}{{\mathrm{CUT}}}

\newcommand{\Cl}{{\mathrm{Cl}}}
\newcommand{\pred}{{\mathrm{pred}}}
\newcommand{\SET}{{\mathrm{SET}}}
\newcommand{\LIST}{{\mathrm{LIST}}}
\newcommand{\mincut}{{\mathrm{mincut}}}
\newcommand{\minord}{{\mathrm{minord}}}
\newcommand{\MinCut}{{\mathrm{MinCut}}}

\SetKw{KwInitialization}{Initialization:}

\begin{document}

\title{Alphabet Size Reduction for Secure Network Coding: A Graph Theoretic Approach}
\author{Xuan~Guang,~\IEEEmembership{Member,~IEEE},
        and~Raymond~W.~Yeung,~\IEEEmembership{Fellow,~IEEE}
        }

\markboth{Alphabet size reduction for secure network coding: a graph theoretic approach}%
{}
{}
%


\maketitle

\begin{abstract}

We consider a communication network where there exist wiretappers who can access a subset of channels, called a {\em wiretap set}, which is chosen from a given collection of wiretap sets. The collection of wiretap sets can be arbitrary. Secure network coding is applied to prevent the source information from being leaked to the wiretappers. In secure network coding, the required alphabet size is an open problem not only of theoretical interest but also of practical importance, because it is closely related to the implementation of such coding schemes in terms of computational complexity and storage requirement.  In this paper, we develop a systematic graph-theoretic approach for improving Cai and Yeung's lower bound on the required alphabet size for the existence of secure network codes. The new lower bound thus obtained, which depends only on the network topology and the collection of wiretap sets, can be significantly smaller than Cai and Yeung's lower bound. A polynomial-time algorithm is devised for efficient computation of the new lower bound.

\end{abstract}
\begin{IEEEkeywords}
Information-theoretic security, secure network coding, wiretap network, alphabet size, lower bound, polynomial-time algorithm, graph theory.
\end{IEEEkeywords}

\IEEEpeerreviewmaketitle

\section{Introduction}

In Shannon's celebrated paper \cite{Shannon-secrecy}, the well-known \textit{Shannon cipher system} is proposed, in which a sender wishes to transmit a private message to a receiver in the presence of a wiretapper, and it is required that the wiretapper can obtain no information about the message. For this purpose, the sender encrypts the message with a random key which is shared with the receiver via a ``secure'' channel and is inaccessible by the wiretapper. The encrypted message is transmitted to the receiver via a ``public'' channel which is eavesdropped by the wiretapper. The receiver can recover the message from the random key and the encrypted message, while the wiretapper obtains no information about the message. In the literature, this is referred to as {\em information-theoretic security}.

Another well-known cipher system of information-theoretic security is \textit{secret sharing}, proposed independently by Blakley \cite{Blakley_secret-sharing-1979} and Shamir \cite{Shamir_secret-sharing-1979}, which is more elaborate than Shannon cipher system. In this system, a secret is encoded into shares which are distributed among a set of participants in such a way that only an arbitrarily specified qualified set of participants can recover the secret, while no information at all about the secret can be obtained from the shares of an unqualified set of participants.

In the context of communications, Ozarow and Wyner \cite{wiretap-channel-II} proposed a related model called \textit{wiretap channel \Rmnum{2}}. In this model, the sender's message is transmitted to the receiver through a set of noiseless point-to-point channels. It is assumed that a wiretapper can fully access any one but not more than one subset of the channels up to a certain size, which is referred to as a wiretap set. Logically, secret sharing contains wiretap channel \Rmnum{2} as a special case.

In 1978, Celebiler and Stette \cite{Celebiler-Stette-1978} proposed a scheme that can improve the efficiency of a two-way satellite communication system by performing the addition of two bits onboard the satellite. In 1999, Yeung and Zhang \cite{Zhang-Yeung-1999} studied the general coding problem in a satellite communication system and obtained an inner bound and an outer bound on the capacity region. In 2000, Ahlswede~\textit{et al.} \cite{Ahlswede-Cai-Li-Yeung-2000} proposed the general concept of network coding that allows the intermediate nodes in a noiseless network to process the received information. In particular, they proved that if coding is applied at the nodes in a network, rather than routing only, the source node can multicast messages to all the sink nodes at the theoretically maximum rate, i.e., the smallest minimum cut capacity between the source node and a sink node, as the alphabet size of both the information source and the channel transmission symbol tends to infinity. Li~\textit{et~al.} \cite{Li-Yeung-Cai-2003} further proved that linear network coding with a finite alphabet is sufficient for optimal multicast by means of a vector space approach. Independently, Koetter and M\'{e}dard \cite{Koetter-Medard-algebraic} developed an algebraic characterization of linear network coding by means of a matrix approach. The above two approaches correspond to the {\em global} and {\em local} descriptions of linear network coding, respectively. Jaggi~\textit{et~al.} \cite{co-construction} proposed a deterministic polynomial-time algorithm for constructing a linear network code. In Tan~\textit{et~al.} \cite{Tan-Yeung-Ho-Cai-Unified-Framework}, the fundamental concept of linear independence among global encoding kernels was studied in depth. Based on this, a unified construction for different classes of linear network codes is obtained.~It~was
shown explicitly in Sun \textit{et al.} \cite{Sun-matroids-LNC} that the linear independence structure of a {\em generic linear network code} naturally induces a matroid. An interesting characterization of the required field size of linear network codes over acyclic multicast networks was recently obtained by Sun~\textit{et~al.}\,\cite{Sun_FieldSizeOnLNC-IT-15}. Their work reveals that the existence of a linear network code over a given finite field does not imply the existence of one over all larger finite fields. For comprehensive discussions of network coding, we refer the reader to \cite{Zhang-book, Yeung-book, Fragouli-book, Fragouli-book-app, Ho-book}.

In the paradigm of network coding, information-theoretic security is naturally considered in the presence of a wiretapper. This problem, called {\em secure network coding}, was introduced by Cai and Yeung in \cite{secure-conference, Cai-Yeung-SNC-IT}. In the wiretap network model of secure network coding, the wiretapper, who can access any one wiretap set of edges, is not allowed to obtain any information about the private source message, while all the sink nodes as legal users can decode the private source message with zero error. Secret sharing can be formulated as a special case of secure network coding.

Similar to the coding for the classical wiretap models \cite{Shannon-secrecy, Blakley_secret-sharing-1979, Shamir_secret-sharing-1979, wiretap-channel-II}, in secure network coding, it is necessary to randomize the source message to guarantee information-theoretic security. El~Rouayheb~\textit{et~al.}~\cite{Rouayheb-IT} showed that the construction of secure network codes in \cite{secure-conference, Cai-Yeung-SNC-IT} can be viewed as a network generalization of the code construction for wiretap channel \Rmnum{2} in \cite{wiretap-channel-II}. Motivated by El~Rouayheb~\textit{et al.}, Silva and Kschischang \cite{Silva-UniversalSNC} proposed a universal design of secure network codes via rank-metric codes such that the design of linear network codes for message transmission and the design of coding for security can be separated.

For secure network coding, the existing bound on the required alphabet size in \cite{Cai-Yeung-SNC-IT, Rouayheb-IT, Silva-UniversalSNC} is roughly equal to the number of all wiretap sets, which is typically too large for implementation in terms of computational complexity and storage requirement. Therefore, the required alphabet size is a problem not only of theoretical interest but also of practical importance. Feldman \textit{et al.} \cite{Feldman} showed that for a given security level, the alphabet size can be reduced by sacrificing a small fraction of the information rate. However, if the information rate is not sacrificed, even for the special case of an {\em $r$-wiretap network}, i.e., the wiretapper can access any one subset of at most $r$ edges, whether it is possible to reduce the required alphabet size is not known \cite{Rouayheb-IT}. Recently, for this special case, Guang \textit{et al.} \cite{Guang-SmlFieldSize-SNC-comm-lett} proposed an equivalence relation of wiretap sets that can be applied to obtain an improved lower bound on the required field size. However, they did not provide any efficient algorithm for computing this bound.

In this paper, we fully explore the underlying mathematical structure of the approach in \cite{Guang-SmlFieldSize-SNC-comm-lett} and show that the required alphabet size for the existence of secure network codes can be reduced significantly, where the collection of the wiretap sets considered here is arbitrary. The main contributions and organization of the paper are given as follows:
\begin{itemize}
\item In Section~\ref{section_preliminaries}, we present secure network coding and the preliminaries, and introduce the necessary notation and definitions.
  \item In Section~\ref{section_bounds}, we generalize the equivalence relation amongst the wiretap sets in $r$-wiretap networks in \cite{Guang-SmlFieldSize-SNC-comm-lett} to general wiretap networks and introduce a domination relation amongst the equivalence classes. We further prove that this domination relation is a strict partial order so that the set of the equivalence classes constitutes a strictly partially ordered set. The number of the maximal elements in this strictly partially ordered set is proved to be a lower bound on the required alphabet size, which in general is a significant improvement over the existing results. Our lower bound is applicable to both linear and non-linear secure network codes, and its improvement over the existing results can be unbounded.
  \item Our lower bound is graph-theoretical, and it depends only on the network topology and the collection of the wiretap sets. Section~\ref{section_algorithms} is devoted to the development of an efficient computation of our lower bound. Toward this end, we introduce the concept of {\em primary minimum cut}, by which we can bypass the complicated operations for determining the equivalence classes of wiretap sets and the domination relation among them. With this, a polynomial-time algorithm is developed for computing the lower bound.
  \item We conclude in Section~\ref{section_conclusion} with a summary of our results and a remark on future research.
\end{itemize}


\section{Preliminaries}\label{section_preliminaries}

In this section, we first present the model of a wiretap network \cite{secure-conference, Cai-Yeung-SNC-IT} to be discussed in this paper. Let $G=(V, E)$ be a finite directed acyclic network with a single source node $s$ and a set of sink nodes $T\subset V\setminus \{s\}$, where $V$ and $E$ are the sets of nodes and edges, respectively. In $G$, let $e=(u,v)\in E$ stand for a directed edge from node $u$ to node $v$, where node $u$ is called the {\em tail} of $e$ and node $v$ is called the {\em head} of $e$, denoted by $\tail(e)$ and $\head(e)$, respectively. Further, for a node $v$, define $\In(v)$ as the set of incoming edges of $v$ and $\Out(v)$ as the set of outgoing edges of $v$. Formally, $\In(v)=\{e \in E:\ \head(e)=v\}$ and $\Out(v)=\{e\in E:\ \tail(e)=v\}$. Without loss of generality, assume $\In(s)=\emptyset$ and $\Out(t)=\emptyset$ for any sink node $t\in T$. An index taken from an alphabet can be transmitted on each edge $e$ in $E$ and parallel edges between two adjacent nodes are allowed. In other words, the capacity of each edge is taken to be $1$. We make this assumption throughput the paper. Let $\mA$ be a collection of subsets of $E$, where every edge set in $\mA$ is called a \textit{wiretap set}. Then a \textit{wiretap network} is specified by a quadruple $(G, s, T, \mA)$, where the source node $s$ generates a source message and injects it into the network; each sink node $t\in T$ as a legal user is required to recover the source message with zero error; arbitrary one wiretap set in $\mA$, but no more than one, may be fully accessed by a wiretapper. The collection $\mA$ of the wiretap sets is known by the source node and sink nodes but which wiretap set in $\mA$ is actually eavesdropped is unknown. Since the source node $s$ and the sink node set $T$ are usually fixed, we use $(G, \mA)$ to denote such a wiretap network for simplicity.

In a network $G$, if a sequence of edges $(e_1,e_2,\cdots,e_m)$ satisfies $\tail(e_1)=u$, $\head(e_m)=v$, and $\tail(e_{k+1})=\head(e_k)$ for all $k=1,2,\cdots,m-1$, we say that the sequence $(e_1,e_2,\cdots,e_m)$ is a path from node $u$ (or edge $e_1$) to node $v$ (or edge $e_m$). A {\em cut} between the source node $s$ and a non-source node $t$ is defined as a set of edges whose removal disconnects $s$ from $t$. The \textit{capacity} of a cut between $s$ and $t$ is defined as the number of edges in the cut, and the minimum of the capacities of all the cuts between $s$ and $t$ is called the \textit{minimum cut capacity} between them. A cut between $s$ and $t$ is called a \textit{minimum cut} if its capacity achieves the minimum cut capacity between them. These concepts can be extended to edge subsets of $E$. We first consider a cut between $s$ and a set of non-source nodes $T$ in the network $G$ as follows. We create a new node $t_T$, and for every node $t$ in $T$, add a new ``super-edge'' of infinite capacity \footnote{Infinite symbols in the alphabet can be transmitted by one use of the edge.} from $t$ to $t_T$ (which is equivalent to adding infinite parallel edges from $t$ to $t_T$). A cut of the finite capacity between $s$ and $t_T$ is defined as a {\em cut} between $s$ and $T$. We can naturally extend the capacity of a cut, the minimum cut capacity and the minimum cut to the case of $T$. Furthermore, let $A\subset E$ be an edge subset. Introduce a node $t_e$ for each edge $e\in A$ which splits $e$ into two edges $e^1$ and $e^2$ with $\tail(e^1)=\tail(e)$, $\head(e^2)=\head(e)$, and $\head(e^1)=\tail(e^2)=t_e$. Let $T_A=\{t_e: e\in A \}$ and then a cut between $s$ and $T_A$ is defined as a {\em cut} between $s$ and $A$. In particular, if $e^1$ or $e^2$ appears in the cut, replace it by $e$. Similarly, the \textit{minimum cut capacity} between $s$ and $A$, denoted by $\mincut(s, A)$, is defined as the minimum cut capacity between $s$ and $T_A$, and a cut between $s$ and $A$ achieving the minimum cut capacity $\mincut(s, A)$ is called a \textit{minimum cut}. If an edge set $B\subseteq E$ is a cut between the source node $s$ and a non-source node $t$ (resp. a set of non-source nodes $T$ and a set of edges $A$), then we say that the edge set $B$ {\em separates} $t$ (resp. $T$ and $A$) from $s$. Note that if $B$ separates $t$ (resp. $T$ and $A$) from $s$, then every path from $s$ to $t$ (resp. $T$ and $A$) passes through at least one edge in $B$.

The following Menger's theorem shows that the minimum cut capacity between node $s$ to node $t$ (resp. $T$ and $A$) and the maximum number of edge-disjoint paths from $s$ to $t$ (resp. $T$ and $A$) are really alternative ways to address the same issue.

\textbf{Edge Version of Menger's Theorem} (\cite[Theorem 6.7]{Book-NetwFlow} and \cite[Theorem 7.16]{Book-GraphTh-Bondy-Murty}):
The maximum number of edge-disjoint paths from node $s$ to node $t$ equals the minimum cut capacity between node $s$ and node $t$.

In secure network coding, the source node $s$ generates a random source message $M$ according to an arbitrary distribution on a message set $\mM$. The source message $M$ is multicast to every sink node $t\in T$, while being protected from the wiretapper who can access any wiretap set $A$ in $\mA$. Similar to the other information-theoretically secure models, in our wiretap network model, it is necessary to randomize the source message to combat the wiretapper. The randomness available at the source node, called the \textit{key}, is a random variable $K$ that takes values in a set of keys $\mK$ according to the uniform distribution.

Let $\mF$ be an alphabet. An $\mF$-valued secure network code on a wiretap network $(G, \mA)$ consists of a set of local encoding mappings $\{ \phi_e: e\in E \}$ such that for every $e$, $\phi_e$ is a mapping from $\mM\times \mK$ to the alphabet $\mF$ if $e\in \Out(s)$, and is a mapping from $\mF^{|\In(v)|}$ to $\mF$ if $e\in \Out(v)$ for a node $v\in V\setminus\{s\}$. The {\em information rate} of the secure network code is $\log_{|\mF|}|\mM|$.

To facilitate our discussion, let $Y_e$ be the random variable transmitted on the edge $e$ that is a function of the random source message $M$ and the random key $K$. For a subset $A$ of $E$, denote~$(Y_e: e\in A)$~by~$Y_A$.

\begin{defn}\label{defn_security}\em
For a secure network code on the wiretap network $(G, \mA)$, $I(Y_A ; M)=0$ for every wiretap set $A\in \mA$, where $I(Y_A ; M)$ denotes the mutual information between $Y_A$ and $M$.
\end{defn}

The notion of security used in the definition of a secure network code is referred in the literature as {\em information-theoretic security} as oppose to {\em computational security}.

\begin{prop}[{\cite[Theorem 3]{Cai-Yeung-SNC-IT}}]\label{thm_bound_Cai_Yeung}\em
Let $(G,\mA)$ be a wiretap network and $\mF$ be an alphabet with $|\mF|\geq |T|$, the number of sink nodes in $G$. Then there exists an $\mF$-valued secure network code over $(G, \mA)$ provided that $|\mF|>|\mA|$. \footnote{The reason for requiring $|\mF|\geq |T|$ here is to guarantee the existence of a network code on $G$. In general, $|\mA|$ is much larger than~$|T|$.}
\end{prop}

If a wiretap set $A\in \mA$ satisfies $|A|=\mincut(s, A)$, then we say that the wiretap set $A$ is \textit{regular}. Further, if all wiretap sets $A$ in $\mA$ are regular, then we say that the collection of wiretap sets $\mA$ is \textit{regular}. For an arbitrary $\mA$, by Proposition~\ref{thm_bound_Cai_Yeung}, there exists a secure network code if $|\mF|>|\mA|$. Now for each $A$ in $\mA$, replace it by a minimum cut $\CUT_A$ between $s$ and $A$ to form $\mA'$. Observe that the minimum cut $\CUT_A$ is regular since $$|\CUT_A|=\mincut(s, A)\leq \mincut(s, \CUT_A)\leq |\CUT_A|,$$
where the first inequality follows from the fact that each cut separating $\CUT_A$ from $s$ also separates $A$ from $s$, and a secure network code which is secure for the wiretap sets in $\mA'$ is also secure for the wiretap sets in $\mA$. Therefore, with respect to the bound given by Proposition~\ref{thm_bound_Cai_Yeung}, it suffices to consider regular wiretap sets. In the rest of the paper, we assume that all edge sets are regular unless otherwise~specified.

In the following, we recall two concepts about strict and non-strict partial orders, which are used frequently in the paper.

\begin{defn}\label{defn_2_partial_orders}\em
Let $\mathfrak{D}$ be a finite set, and let ``$<$'' and ``$\leq$'' be two binary relations amongst the elements in $\mathfrak{D}$.
\begin{itemize}
  \item The binary relation ``$<$'' is called a strict partial order in $\mathfrak{D}$ if the following conditions are satisfied for arbitrary elements $a$, $b$, and $c$ in $\mathfrak{D}$:
  \begin{enumerate}
     \item {\bf(Irreflexivity)} $a \nless a$;
     \item {\bf(Transitivity)} if $a < b$ and $b < c$, then $a < c$;
     \item {\bf(Asymmetry)} if $a < b$, then $b \nless a$.\footnote{Asymmetry can readily be deduced from irreflexivity and transitivity.}
  \end{enumerate}
  \item The binary relation ``$\leq$'' is called a non-strict partial order in $\mathfrak{D}$ if the following conditions are satisfied for arbitrary elements $a$, $b$, and $c$ in $\mathfrak{D}$:
  \begin{enumerate}
    \item {\bf(Reflexivity)} $a\leq a$;
    \item {\bf(Antisymmetry)} if $a \leq b$ and $b \leq a$, then $a=b$;
    \item {\bf(Transitivity)} if $a \leq b$ and $b \leq c$, then $a \leq c$.
  \end{enumerate}
\end{itemize}
\end{defn}

\section{Required Alphabet Size for Secure Network Coding}\label{section_bounds}

In this section, for a wiretap network $(G, \mA)$, we prove a new bound on the required alphabet size of the existence of secure network codes that improves upon the lower bound in \cite{Guang-SmlFieldSize-SNC-comm-lett}. In the next section, we present an efficient algorithm for evaluating this bound.

Let $A$ and $A'$ be two edge sets in $G$. Define a binary relation ``$\sim$'' between $A$ and $A'$: $A \sim A'$ if and only if there exists an edge set $\CUT$ which is a minimum cut between $s$ and $A$ and also between $s$ and $A'$, that is, $A$ and $A'$ have a common minimum cut between the source node $s$ and each of them. Note that $A \sim A'$ implies $|A|=|A'|$ because $\mincut(s, A)=|\CUT|=\mincut(s, A')$ and both $A$ and $A'$ are regular. It was proved in \cite{Guang-SmlFieldSize-SNC-comm-lett} that ``$\sim$'' is an equivalence relation. While reflexivity and symmetry of ``$\sim$'' are immediate, the proof of transitivity is nontrivial.

With the relation ``$\sim$'', the wiretap sets in $\mA$ can be partitioned into equivalence classes. All the wiretap sets in an equivalence class have a common minimum cut, which is implied by the transitivity  of ``$\sim$''. To see this, consider wiretap sets $A$, $A'$, and $A''$ that are in the same equivalence class. Let $A$ and $A'$ have a common minimum cut $\CUT$, and let $A'$ and $A''$ have a common minimum cut $\CUT'$. Then $\CUT \sim A'$ and $\CUT' \sim A'$. By the transitivity of ``$\sim$'', we have $\CUT \sim \CUT'$, implying that there exists a common minimum cut between $s$ and $\CUT$ and between $s$ and $\CUT'$, which in turn is a minimum cut between $s$ and $A$, between $s$ and $A'$, and between $s$ and $A''$. Then we see by induction that all the wiretap sets in the equivalence class can have a common minimum cut from $s$. Immediately, we give the following proposition.

\begin{prop}\label{lem_1}\em
Let $A_1, A_2, \cdots, A_m$ be $m$ equivalent edge sets under the equivalence relation ``$\sim$''.~Then
\begin{align}\label{equ_equi-union}
\mincut(s, \cup_{i=1}^{m}A_i)=\mincut(s, A_j), \quad \forall j,\ 1\leq j \leq m.
\end{align}
\end{prop}
\begin{IEEEproof}
Since the edge sets $A_1, A_2, \cdots, A_m$ are equivalent, they have a common minimum cut $\CUT$ separating each of them from $s$. This implies that
\begin{enumerate}
  \item $|\CUT|=\mincut(s, A_j)$, $\forall j$, $1 \leq j \leq m$;\label{L1-lem_1}
  \item $\CUT$ is a cut between $s$ and $\cup_{i=1}^{m}A_i$.\label{L2-lem_1}
\end{enumerate}
Combining \ref{L1-lem_1}) and \ref{L2-lem_1}), we have that for all $j$, $1 \leq j \leq m$,
$$|\CUT|=\mincut(s, A_j)\leq \mincut(s, \cup_{i=1}^{m}A_i) \leq |\CUT|,$$
completing the proof.
\end{IEEEproof}

Let $N(\mA)$ be the number of the equivalence classes in $\mA$. According to Proposition~\ref{thm_bound_Cai_Yeung} and the discussions that follow, by replacing each equivalence class of wiretap sets in $\mA$ by its common minimum cut, we see that there exists an $\mF$-valued secure network code on $(G, \mA)$ provided that $|\mF|> N(\mA)$. This lower bound $N(\mA)$ on $|\mF|$ was originally obtained in \cite{Guang-SmlFieldSize-SNC-comm-lett} for $r$-wiretap networks, but it also applies for general wiretap networks. We use the following example to illustrate the advantage of this approach.

\vspace{-3mm}
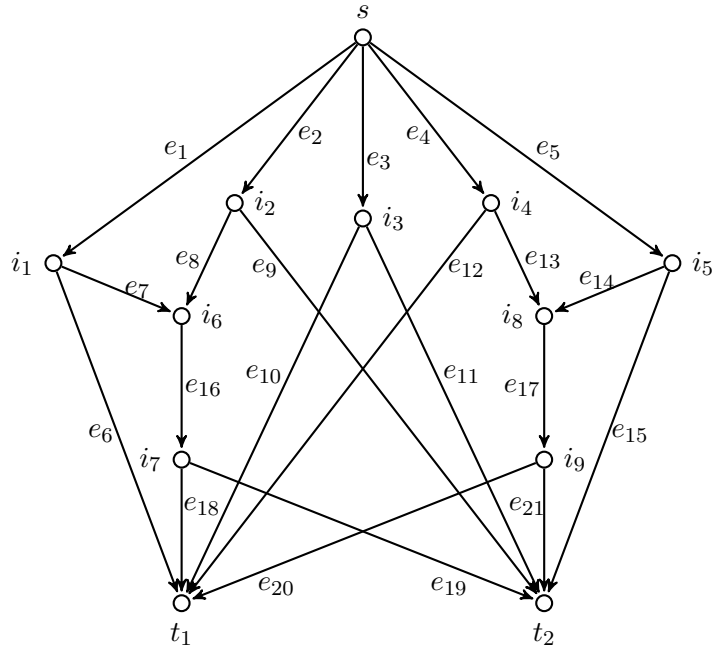
\begin{figure}[!htb]
\centering
\begin{tikzpicture}
[->,>=stealth',shorten >=1pt,auto,node distance=2.41cm, thick]
  \tikzstyle{every state}=[fill=none,draw=black,text=black,minimum size=6pt,inner sep=0pt]
  \node[state]         (s)[label=above:$s$]                 {};
  \node[state]         (i_2)[below left of=s, xshift=0mm, yshift=-5mm, label=right:$i_2$]   {};
  \node[state]         (i_4)[below right of=s, xshift=0mm, yshift=-5mm, label=right:$i_4$] {};
  \node[state]         (i_3)[below of=s, yshift=0mm, label=right:$i_3$] {};
  \node[state]         (i_1)[left of=i_2, xshift=0mm, yshift=-8mm, label=left:$i_1$] {};
  \node[state]         (i_5)[right of=i_4, xshift=0mm, yshift=-8mm, label=right:$i_5$] {};
  \node[state]         (i_6)[below right of=i_1, yshift=10mm, label=right:$i_6$] {};
  \node[state]         (i_7)[below of=i_6, yshift=5mm, label=left:$i_7$] {};
  \node[state]         (i_8)[below left of=i_5, yshift=10mm, label=left:$i_8$] {};
  \node[state]         (i_9)[below of=i_8, yshift=5mm, label=right:$i_9$] {};
  \node[state]         (t_1)[below of=i_7, yshift=5mm, label=below:$t_1$] {};
  \node[state]         (t_2)[below of=i_9, yshift=5mm, label=below:$t_2$] {};
\path
(s) edge           node[left=1mm] {$e_1$} (i_1)
    edge           node[pos=0.6, right=0mm] {$e_2$} (i_2)
    edge           node[pos=0.7, right=-1mm] {$e_3$} (i_3)
    edge           node[pos=0.6, left=-0.5mm] {$e_4$} (i_4)
    edge           node[right=1mm] {$e_5$} (i_5)
(i_1) edge         node[left=-1mm]{$e_6$}(t_1)
      edge       node[pos=0.5, right=-0.5mm]{$e_7$}(i_6)
(i_2) edge       node[pos=0.5, left=-0.5mm]{$e_8$}(i_6)
      edge       node[pos=0.15, left=-0.5mm]{$e_9$}(t_2)
(i_3) edge       node[pos=0.4, left=-0.5mm]{$e_{10}$}(t_1)
      edge       node[pos=0.4, right=-0.5mm]{$e_{11}$}(t_2)
(i_4) edge       node[pos=0.15, right=-0.5mm]{$e_{12}$}(t_1)
      edge       node[pos=0.5, right=-0.5mm]{$e_{13}$}(i_8)
(i_5) edge       node[pos=0.3, left=0.5mm]{$e_{14}$}(i_8)
      edge       node[right=-1mm]{$e_{15}$}(t_2)
(i_6) edge       node[pos=0.5, right=-1mm]{$e_{16}$}(i_7)
(i_8) edge       node[pos=0.5, left=-1mm]{$e_{17}$}(i_9)
(i_7) edge       node[pos=0.3, right=-1.2mm]{$e_{18}$}(t_1)
      edge       node[pos=0.9, left=3mm]{$e_{19}$}(t_2)
(i_9) edge       node[pos=0.9, right=3mm]{$e_{20}$}(t_1)
      edge       node[pos=0.3, left=-1.7mm]{$e_{21}$}(t_2);
\end{tikzpicture}
\vspace{-3mm}
\caption{The network $G$.}
\label{Fig_CombNet-4-3}
\end{figure}

\begin{eg}\label{eg_Ordered-Set-1}\em
Consider the network $G$ depicted in Fig.~\ref{Fig_CombNet-4-3}.
Let the collection of wiretap sets $\mA$ be
\begin{align*}
\mA=&\Big\{ \{ e_6 \}, \{ e_7 \}, \{ e_8 \}, \{ e_9 \}, \{ e_{12} \}, \{ e_{13} \}, \{ e_{14} \}, \{ e_{15} \}, \{ e_{18} \}, \{ e_{19} \}, \{ e_{20} \},\{ e_{21} \},\\
    &\{ e_{6}, e_{18} \}, \{ e_{6}, e_{19} \}, \{ e_{7}, e_{18} \}, \{ e_{7}, e_{19} \},
     \{ e_{8}, e_{11} \}, \{ e_{8}, e_{16} \}, \{ e_{8}, e_{18} \}, \{ e_{9}, e_{10} \},\\
    &\{ e_{9}, e_{18} \}, \{ e_{9}, e_{19} \}, \{ e_{10}, e_{14}\}, \{ e_{10}, e_{15}\},
     \{ e_{10}, e_{19}\}, \{ e_{10}, e_{21}\}, \{ e_{11}, e_{14}\}, \{ e_{11}, e_{15}\},\\
    &\{ e_{11}, e_{18}\}, \{ e_{11}, e_{20}\}, \{ e_{12}, e_{20}\}, \{ e_{12}, e_{21}\},
     \{ e_{13}, e_{17}\}, \{ e_{13}, e_{21}\}, \{ e_{14}, e_{20}\}, \{ e_{14}, e_{21}\},\\
    &\{ e_{15}, e_{20}\}, \{ e_{15}, e_{21}\}, \{ e_{18}, e_{20}\}, \{ e_{18}, e_{21}\},
     \{ e_{19}, e_{20}\}, \{ e_{19}, e_{21}\},\\
    &\{ e_{1}, e_{3}, e_{16} \}, \{ e_{1}, e_{11}, e_{16} \}, \{ e_{2}, e_{10}, e_{16} \}, \{ e_{3}, e_{5}, e_{17} \}, \{ e_{4}, e_{10}, e_{17} \}, \{ e_{5}, e_{11}, e_{17} \} \Big\},
\end{align*}
with $|\mA|=48$. The equivalence classes of wiretap sets are
\begin{align*}
&\Cl_1=\Big\{ \{ e_6 \}, \{ e_7 \} \Big\}, \qquad \Cl_2=\Big\{ \{ e_8 \}, \{ e_9 \} \Big\}, \qquad \Cl_3=\Big\{ \{ e_{12} \},\{ e_{13} \}\Big\},\\
&\Cl_4=\Big\{ \{ e_{14} \},\{ e_{15} \}\Big\}, \quad \Cl_{5}=\Big\{ \{ e_{18} \}, \{ e_{19} \} \Big\}, \quad  \Cl_{6}=\Big\{ \{ e_{20} \},\{ e_{21} \}\Big\},\\
&\Cl_{7}=\Big\{ \{ e_{8}, e_{11} \}, \{ e_{9}, e_{10} \} \Big\}, \quad
 \Cl_{8}=\Big\{ \{ e_{10}, e_{19} \}, \{ e_{11}, e_{18} \} \Big\},\quad
 \Cl_{9}=\Big\{ \{ e_{10}, e_{21} \}, \{ e_{11}, e_{20} \} \Big\},\\
&\Cl_{10}=\Big\{ \{ e_{10}, e_{14} \}, \{ e_{10}, e_{15} \}, \{ e_{11}, e_{14} \},\{ e_{11}, e_{15} \} \Big\},\\
&\Cl_{11}=\Big\{ \{ e_{18}, e_{20} \}, \{ e_{18}, e_{21} \}, \{ e_{19}, e_{20} \},\{ e_{19}, e_{21} \} \Big\},\\
&\Cl_{12}=\Big\{ \{ e_{6}, e_{18} \}, \{ e_{6}, e_{19} \}, \{ e_{7}, e_{18} \},\{ e_{7}, e_{19} \},\{ e_{8}, e_{16} \},\{ e_{8}, e_{18} \},\{ e_{9}, e_{18} \}, \{ e_{9}, e_{19} \} \Big\},\\
&\Cl_{13}=\Big\{ \{ e_{12}, e_{20} \}, \{ e_{12}, e_{21} \}, \{ e_{13}, e_{17} \},\{ e_{13}, e_{21} \},\{ e_{14}, e_{20} \},\{ e_{14}, e_{21} \},\{ e_{15}, e_{20} \}, \{ e_{15}, e_{21} \} \Big\},\\
&\Cl_{14}=\Big\{ \{ e_{1}, e_{3}, e_{16} \}, \{ e_{1}, e_{11}, e_{16} \}, \{ e_{2}, e_{10}, e_{16} \} \Big\},\\
&\Cl_{15}=\Big\{ \{ e_{3}, e_{5}, e_{17} \}, \{ e_{4}, e_{10}, e_{17} \}, \{ e_{5}, e_{11}, e_{17} \} \Big\}.
\end{align*}
Then $N(\mA)=15$, which is considerably smaller than $|\mA|$.
\end{eg}

However, \cite{Guang-SmlFieldSize-SNC-comm-lett} does not provide an algorithm for computing $N(\mA)$, making the bound practically not useful except for very simple networks for which $N(\mA)$ can be readily evaluated. This issue will be addressed in the next section after we have introduced the notion of equivalence-class domination in the rest of this section.

The equivalence class containing a wiretap set $A$ is denoted by $\Cl(A)$, or simply $\Cl$ if there is no ambiguity. Note that the wiretap sets have possibly different cardinalities, and a wiretap set may be separated from $s$ by another wiretap set of a larger cardinality. If every wiretap set in an equivalence class can be separated by some wiretap set with a larger cardinality, then it is not necessary to consider this equivalence class for the purpose of lower bounding the required alphabet size. For instance in Example~\ref{eg_Ordered-Set-1}, since both the wiretap sets $\{ e_{18} \}$ and $\{ e_{19} \}$ are separated by another wiretap set $\{ e_{1}, e_{3},e_{16} \}$, it is not necessary to consider $\Cl_{5}=\Big\{ \{ e_{18} \}, \{ e_{19} \} \Big\}$. In the following, we explore the essence of this observation and establish in Theorem~\ref{thm_partial-order-domination} a strict partial order amongst the equivalence classes, which can help further reduce the required alphabet size.

\begin{defn}[\textbf{Wiretap-Set Domination}]\label{defn_domination}\em
Let $A_1$ and $A_2$ be two wiretap sets in $\mA$ with $|A_1|<|A_2|$. We say that $A_1$ is dominated by $A_2$, denoted by $A_1\prec A_2$, if there exists a minimum cut between $s$ and $A_2$ that also separates $A_1$ from $s$. In other words, upon deleting the edges in the minimum cut between $s$ and $A_2$, $s$ and $A_1$ are also disconnected.
\end{defn}

Note that in the above definition, in order for $A_1 \prec A_2$, $|A_1|$ has to be {\em strictly smaller} than $|A_2|$, and $A_1 \prec A_2$ does not mean that $A_2$ is at the ``upstream'' of $A_1$. For instance in Fig.~\ref{Fig_CombNet-4-3}, let $A_1=\{e_3, e_8 \}$ and $A_2=\{e_6, e_{10}, e_{18} \}$. We have $A_2 \succ A_1$ since $\{e_1, e_{2}, e_{3}\}$ is a minimum cut between $s$ and $A_2$ that separates $A_1$ from $s$, although $A_1$ is actually at the ``upstream'' of $A_2$.

The following proposition gives a necessary and sufficient condition for the existence of a domination relation between two wiretap sets.

\begin{prop}\label{prop_wiretap-set}\em
For wiretap sets $A_1$ and $A_2$ such that $|A_1|<|A_2|$, $A_1\prec A_2$ if and only if
\begin{align}\label{equ_in_prop_wiretap-set}
\mincut(s, A_1\cup A_2)=\mincut(s, A_2).
\end{align}
\end{prop}
\begin{IEEEproof}
By Definition~\ref{defn_domination}, the ``only if'' part is evident. We only need to prove the ``if'' part. Let $\CUT$ be a minimum cut separating $A_1\cup A_2$ from $s$, so that by (\ref{equ_in_prop_wiretap-set}),
\[|\CUT|=\mincut(s, A_1\cup A_2)=\mincut(s, A_2),\]
which implies that $\CUT$ is also a minimum cut between $s$ and $A_2$. Since $\CUT$ is also a cut (not minimum because $|A_1|<|A_2|$) between $s$ and $A_1$, we have $A_1\prec A_2$ by definition. This completes the proof.
\end{IEEEproof}

We remark that although (\ref{equ_equi-union}) for $m=2$ is equivalent to (\ref{equ_in_prop_wiretap-set}), Proposition~\ref{lem_1} and Proposition~\ref{prop_wiretap-set} are different because in Proposition~\ref{lem_1}, $A_1, A_2, \cdots, A_m$ have the same cardinality, while  in Proposition~\ref{prop_wiretap-set} we have $|A_1|<|A_2|$.

Next, we extend the notion of domination to equivalence classes.

\begin{defn}[\textbf{Equivalence-Class Domination}]\label{defn_cl-domin}\em
For two distinct equivalence classes $\Cl_1$ and $\Cl_2$, if there exists a common minimum cut of the wiretap sets in $\Cl_2$ that separates all the wiretap sets in $\Cl_1$ from $s$, we say that $\Cl_1$ is dominated by $\Cl_2$, denoted by $\Cl_1 \prec \Cl_2$.\footnote{Here we use the same symbol ``$\prec$'' to represent two domination relations, but this abuse of notation should cause no ambiguity.}
\end{defn}

We also give a necessary and sufficient condition for the existence of a domination relation between two equivalence classes.

\begin{thm}\label{thm_cl-domination}\em
Let $A_1$ and $A_2$ be two wiretap sets in $\mA$. Then $\Cl(A_1)\prec \Cl(A_2)$ if and only if $A_1'\prec A_2'$ for all $A_1' \in \Cl(A_1)$ and $A_2' \in \Cl(A_2)$.
\end{thm}
\begin{IEEEproof}
See Appendix~\ref{app:a}.
\end{IEEEproof}

For the equivalence-class domination relation, we have the following theorem.

\begin{thm}\label{thm_partial-order-domination}\em
The equivalence-class domination relation ``$\prec$'' amongst the equivalence classes in $\mA$ is a strict partial order.
\end{thm}

In order to prove Theorem~\ref{thm_partial-order-domination}, we need the following lemma. Denote by $\MinCut(B)$ the set of the minimum cuts between $s$ and an edge set $B$.

\begin{lemma}\label{lem_2}\em
Let $A_1$ and $A_2$ be two wiretap sets and $A_1\prec A_2$. Then for any $\CUT_1\in \MinCut(A_1)$ and any $\CUT_{1,2} \in \MinCut(A_1\cup A_2)$, $$\CUT_{1}\prec \CUT_{1,2}.$$
\end{lemma}
\begin{IEEEproof}
See Appendix~\ref{app:b}.
\end{IEEEproof}

An important consequence of Lemma~\ref{lem_2} is the following theorem which enhances Theorem~\ref{thm_cl-domination}.

\begin{thm}\label{cor_1}\em
$\Cl(A_1)\prec \Cl(A_2)$ if and only if $A_1\prec A_2$.
\end{thm}
\begin{IEEEproof}
By Definition~\ref{defn_cl-domin}, the ``only if'' part is obvious. In the following we prove the ``if'' part. Let $\CUT_i$ be a common minimum cut of the wiretap sets in $\Cl(A_i)$, $i=1,2$. Let $\CUT_{1,2} \in \MinCut(A_1\cup A_2)$, and so $\CUT_{1,2}\sim A_2$ by Proposition~\ref{prop_wiretap-set}.

Since $A_1\prec A_2$,  by Lemma~\ref{lem_2} we have $\CUT_1\prec \CUT_{1,2}$, i.e., there exists a minimum cut $\CUT$ between $s$ and $\CUT_{1,2}$ that separates $\CUT_1$ from $s$. By Proposition~\ref{prop_wiretap-set}, we further obtain
$$\mincut(s, \CUT_1 \cup \CUT_{1,2})=\mincut(s, \CUT_{1,2}),$$
implying that $\CUT \sim \CUT_{1,2} \sim A_2$. Together with $A_2 \sim \CUT_2$, we obtain $\CUT \sim \CUT_2$. Thus, $\CUT$ and $\CUT_2$ have a common minimum cut, denoted by $\CUT^*$, which satisfies the following:
\begin{enumerate}
  \item $\CUT^*$ is a common minimum cut between $s$ and each of the wiretap sets in $\Cl(A_2)$, since $\CUT^*$ is a minimum cut between $s$ and $\CUT_2$.
  \item $\CUT^*$ separates each of the wiretap sets in $\Cl(A_1)$, since $\CUT^*$ is a minimum cut between $s$ and $\CUT$, and $\CUT$ separates $\CUT_1$ from $s$.
\end{enumerate}
It then follows by definition that $\Cl(A_1)\prec \Cl(A_2)$, completing the proof.
\end{IEEEproof}

With lemma~\ref{lem_2} and Theorem~\ref{cor_1}, we are now ready to prove Theorem~\ref{thm_partial-order-domination}.

\begin{IEEEproof}[Proof of Theorem~\ref{thm_partial-order-domination}]
The irreflexivity can be easily proved by Definition~\ref{defn_domination} and Theorem~\ref{cor_1} as follows. Assume $\Cl \prec \Cl$ for some equivalence class $\Cl$, which implies by Theorem~\ref{cor_1} that $A\prec A$ for any $A\in \Cl$, a contradiction to the definition of wiretap-set domination (Definition~\ref{defn_domination}).

To complete the proof, we only need to prove the transitivity of ``$\prec$'', i.e., for three equivalence classes $\Cl_1$, $\Cl_2$, and $\Cl_3$, if $\Cl_1 \prec \Cl_2$ and $\Cl_2 \prec \Cl_3$, then $\Cl_1 \prec \Cl_3$. Let $A_i\in \Cl_i$, $i=1,2,3$. By Theorem~\ref{cor_1} and Proposition~\ref{prop_wiretap-set}, it is sufficient to prove that
\begin{align}\label{equ_Thm5}
\mincut(s, A_1\cup A_3)=\mincut(s, A_3).
\end{align}

First, note that a cut separating $A_1\cup A_3$ from $s$ is also a cut between $s$ and $A_3$, which implies
\begin{align}\label{equ1-thm_partial-order-domination}
\mincut(s, A_1\cup A_3) \geq \mincut(s, A_3).
\end{align}
On the other hand, in light of $\Cl_1 \prec \Cl_2$ and $\Cl_2 \prec \Cl_3$, by Definition~\ref{defn_cl-domin}, there exists a common minimum cut $\CUT'$ of the wiretap sets in $\Cl_2$ which separates all the wiretap sets in $\Cl_1$ from $s$, and there exists a common minimum cut $\CUT''$ of the wiretap sets in $\Cl_3$ which separates all the wiretap sets in $\Cl_2$. Consequently, we have $\CUT'\in \MinCut(A_2)$ and $\CUT''\in\MinCut(A_2\cup A_3)$. In addition, we also have $A_2\prec A_3$ by Theorem~\ref{cor_1} since $\Cl_2 \prec \Cl_3$. Thus, it follows from Lemma~\ref{lem_2} that
\begin{align}\label{equ12}
\CUT' \prec \CUT''.
\end{align}

Consider $\CUT'\cup \CUT''$, and note that a cut between $s$ and $\CUT'\cup \CUT''$ separates $A_1\cup A_2 \cup A_3$ from $s$. This implies
\begin{align}\label{equ2-thm_partial-order-domination}
\mincut(s, A_1\cup A_3)\leq \mincut(s, A_1\cup A_2\cup A_3)\leq \mincut(s, \CUT'\cup \CUT'').
\end{align}
Together with (\ref{equ12}) and Proposition~\ref{prop_wiretap-set}, we further obtain that
\begin{align}\label{equ3-thm_partial-order-domination}
\mincut(s, \CUT'\cup \CUT'')=\mincut(s, \CUT'').
\end{align}
Since $\CUT''$ is a common minimum cut of the wiretap sets in $\Cl_3$, we have
\begin{align}
\mincut(s, \CUT'')=\mincut(s, A_3).\label{equ11}
\end{align}
By \eqref{equ2-thm_partial-order-domination}, \eqref{equ3-thm_partial-order-domination}, and \eqref{equ11}, we obtain
\begin{align}
\mincut(s, A_1\cup A_3)\leq \mincut(s, A_3).\label{equ10}
\end{align}
Then \eqref{equ_Thm5} follows from \eqref{equ1-thm_partial-order-domination} and \eqref{equ10}. The theorem is proved.
\end{IEEEproof}

Since the set of all the equivalence classes in $\mA$ has been proved to be a strictly partially ordered set, we can define its \textit{maximal equivalence classes} as follows.

\begin{defn}[\textbf{Maximal Equivalence Class\footnotemark}]\em
For a collection of wiretap sets $\mA$, an equivalence class $\Cl$ is a maximal equivalence class if there exists no other equivalence class $\Cl'$ such that $\Cl'\succ \Cl$. Denote by $N_{\max}(\mA)$ the number of the maximal equivalence classes in $\mA$.
\end{defn}
\footnotetext{The maximal equivalence classes are those maximal elements in the set of equivalence classes, when this set is viewed as a strictly partially ordered set.}

Let $\Cl$ be a maximal equivalence class and $\Cl_1$, $\Cl_2$, $\cdots$, $\Cl_m$ be $m$ equivalence classes that are dominated by $\Cl$. By Definition~\ref{defn_cl-domin}, for each $1\leq i \leq m$, there exists a common minimum cut of the wiretap sets in $\Cl$, denoted by $\CUT_i$, that separates all the wiretap sets in $\Cl_i$ from $s$. For any wiretap set $A$ in $\Cl$, since each $\CUT_i$ is a common minimum cut of the wiretap sets in $\Cl$, $\CUT_i \sim A$. This implies that all $\CUT_i$, $1\leq i \leq m$, are equivalent by transitivity of the equivalence relation ``$\sim$''. Using the argument immediately above Proposition~\ref{lem_1}, we see that $\CUT_i$, $1\leq i \leq m$, have a common minimum cut, say $\CUT$. Then a secure network code which is secure for $\CUT$ is also secure for all the wiretap sets in every $\Cl_i$, $1\leq i \leq m$. Therefore, the number of maximal equivalence classes in $\mA$ gives a new lower bound on the required alphabet size, which is potentially an improvement over the lower bound~$N(\mA)$.

\begin{thm}\label{thm_lower_bound_N_max}\em
Let $(G,\mA)$ be a wiretap network and $\mF$ be the alphabet with $|\mF|\geq |T|$, the number of sink nodes in $G$. Then there exists an $\mF$-valued secure network code on $(G,\mA)$ provided that the alphabet size $|\mF|>N_{\max}(\mA)$.
\end{thm}

We continue to use the setup in Example~\ref{eg_Ordered-Set-1} to illustrate the concepts mentioned above and the advantage of the new bound.

\begin{eg}\label{eg_Ordered-Set-2}\em
Recall the wiretap network $(G, \mA)$ in Example~\ref{eg_Ordered-Set-1}. With the equivalence-class domination ``$\prec$'', the strict partial order of the equivalence classes is illustrated by the Hasse diagram in Fig.~\ref{Fig_order-set}, which shows that $\Cl_{11}$, $\Cl_{14}$, and $\Cl_{15}$ are all of the maximal equivalence classes, i.e., $N_{\max}(\mA)=3$, which is much smaller than $N(\mA)=15$.

\begin{figure}[!htb]
\centering
\begin{tikzpicture}
[latex-, >=stealth', shorten >=1pt, auto, node distance=2.4cm, ultra thick,
	state/.style =
	{
		draw			= black,
		thick,
		ellipse,
		fill			= gray!20,
		minimum height	= 2em
	}
]
  \node[state]         (Cl_{1,2})                 {\footnotesize$\Cl_{12}$};
  \node[state]         (Cl_{2,3})[right of=Cl_{1,2}]   {\footnotesize$\Cl_{7}$};
  \node[state]         (Cl_{4,5})[right of=Cl_{2,3}]   {\footnotesize$\Cl_{13}$};
  \node[state]         (Cl_{3,5})[right of=Cl_{4,5}]   {\footnotesize$\Cl_{10}$};
  \node[state]         (Cl_{3,16})[right of=Cl_{3,5}]   {\footnotesize$\Cl_{8}$};
  \node[state]         (Cl_{3,17})[right of=Cl_{3,16}]   {\footnotesize$\Cl_{9}$};
  \node[state]         (Cl_{16,17})[right of=Cl_{3,17}]   {\footnotesize$\Cl_{11}$};
  \node[state]         (Cl_{1,2,3})[above of=Cl_{2,3}]   {\footnotesize$\Cl_{14}$};
  \node[state]         (Cl_{3,4,5})[above of=Cl_{3,5}]   {\footnotesize$\Cl_{15}$};
  \node[state]         (Cl_{1})[below of=Cl_{1,2}]   {\footnotesize$\Cl_{1}$};
  \node[state]         (Cl_{2})[right of=Cl_{1}]   {\footnotesize$\Cl_{2}$};
  \node[state]         (Cl_{4})[right of=Cl_{2}]   {\footnotesize$\Cl_{3}$};
  \node[state]         (Cl_{5})[right of=Cl_{4}]   {\footnotesize$\Cl_{4}$};
  \node[state]         (Cl_{16})[right of=Cl_{5}]   {\footnotesize$\Cl_{5}$};
  \node[state]         (Cl_{17})[right of=Cl_{16}]   {\footnotesize$\Cl_{6}$};

  \path
(Cl_{1,2,3}) edge           node{}  (Cl_{1,2})
             edge           node{}  (Cl_{2,3})
             edge           node{}  (Cl_{3,16})
(Cl_{3,4,5}) edge           node{}  (Cl_{4,5})
             edge           node{}  (Cl_{3,5})
             edge           node{}  (Cl_{3,17})
(Cl_{1,2})   edge           node{}  (Cl_{1})
             edge           node{}  (Cl_{2})
(Cl_{2,3})   edge           node{}  (Cl_{2})
(Cl_{4,5})   edge           node{}  (Cl_{4})
             edge           node{}  (Cl_{5})
(Cl_{3,5})   edge           node{}  (Cl_{5})
(Cl_{3,16})  edge           node{}  (Cl_{16})
(Cl_{3,17})  edge           node{}  (Cl_{17})
(Cl_{16,17}) edge           node{}  (Cl_{16})
             edge           node{}  (Cl_{17});
\end{tikzpicture}
\caption{The Hasse diagram of the set of all $15$ equivalence  classes, ordered by the equivalence-class domination relation ``$\prec$''.}
\label{Fig_order-set}
\end{figure}
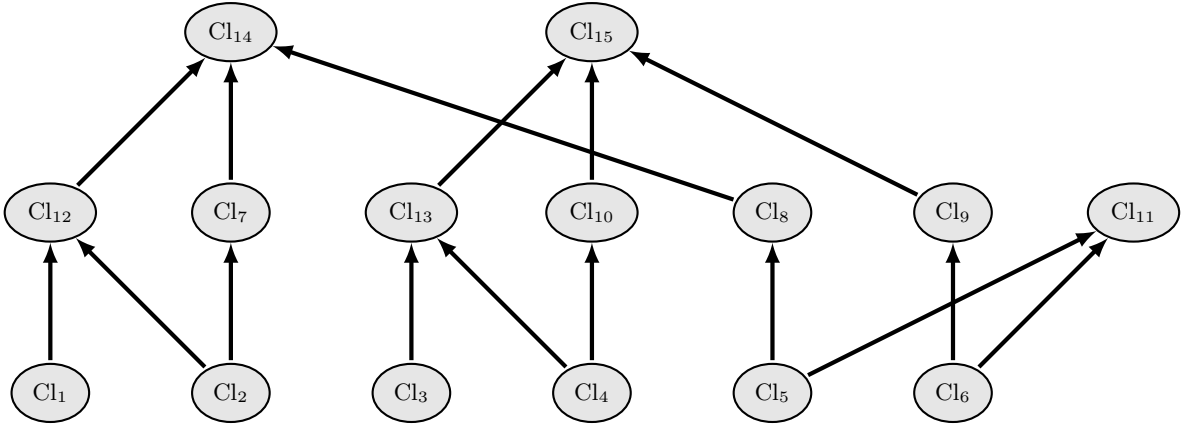

In general, computing the values of $N(\mA)$ and $N_{\max}(\mA)$, or characterizing the corresponding Hasse diagram, is nontrivial. Even in the simple example, their values are not obvious. How to efficiently compute $N(\mA)$ and $N_{\max}(\mA)$ will be discussed in the next section.
\end{eg}

It is easily seen that $N_{\max}(\mA)\leq N(\mA)\leq |\mA|$, and in general $N_{\max}(\mA)$ can be much smaller than $|\mA|$ as illustrated by Example~\ref{eg_Bound-1} below. The only case when $N_{\max}(\mA)$ has no improvement over $|\mA|$, i.e., $N_{\max}(\mA)=|\mA|$, is that every wiretap set itself forms an equivalence class and no domination relation exists amongst all the equivalence classes. In this case, the collection of wiretap sets $\mA$ is ``sparse'' and the value of $|\mA|$ is already small.

\begin{eg}\label{eg_Bound-1}\em
Consider the combination network $G_{N,k}$ (see \cite[p.26]{Zhang-book}, \cite[p.450]{Yeung-book}) with $N=20$ and $k=19$. In this network, there is a single source node $s$, $20$ intermediate nodes, and $|T|={20 \choose 19}=20$ sink nodes. Each intermediate node is connected to $s$, and each sink node is connected to a distinct subset of $19$ intermediate nodes. The number of the edges in the network is $|E|=N+|T|\cdot k=400$, and the minimum cut capacity between $s$ and every sink node is $19$.

We partition all the edges into two layers: the upper and lower layers. The upper layer consists of $N=20$ edges connecting the source node $s$ and the intermediate nodes. The lower layer consists of $|T|\cdot k=380$ edges connecting the intermediate nodes and the sink nodes. Assume that a wiretapper eavesdrops none of the edges in the upper layer and at most $17$ edges in the lower layer, all of which are from distinct intermediate nodes. Note that the number of outgoing edges of each intermediate node is ${N-1 \choose k-1}=19$. Then the total number of wiretap sets is
\begin{align*}
|\mA|=\sum_{i=1}^{17} {20 \choose i}19^i \gg 19^{17} \approx 5.48\times 10^{21}.
\end{align*}

Next, we compute $N(\mA)$ and $N_{\max}(\mA)$. Note that two wiretap sets $A_1$ and $A_2$ with $|A_1|=|A_2|$ are equivalent if and only if
$$\{ \tail(d):\ d\in A_1 \}=\{ \tail(e):\ e\in A_2 \}.$$
Then both $A_1$ and $A_2$ are dominated by the edge subset $\{ d'=(s, \tail(d)):\ d\in A_1 \}$. In general, for every wiretap set $A$ in an equivalence class $\Cl$, $\{ \tail(e):\ e\in A \}$ are identical, and all the wiretap sets in $\Cl$ are dominated by the edge subset $\{e'=(s, \tail(e)):\ e\in A \}$. Thus, there is a one-to-one correspondence between an equivalence class and a subset of intermediate nodes, which has cardinality no more than $17$. Then $N(\mA)=\sum_{i=1}^{17} {20 \choose i}\approx 1.05\times 10^6$, which is much smaller than $|\mA|$.

Furthermore, an equivalence class $\Cl_1$ is dominated by another one $\Cl_2$ if and only if
$\{ \tail(d):\ d\in A_1 \}\subsetneq\{ \tail(e):\ e\in A_2 \}$, where $A_1$ and $A_2$ are two arbitrary wiretap sets in $\Cl_1$ and $\Cl_2$, respectively. In other words, $\Cl_1 \prec \Cl_2$ if and only if the subset of intermediate nodes corresponding to $\Cl_1$ is strictly contained by the subset of intermediate nodes corresponding to $\Cl_2$. Thus, $N_{\max}(\mA)={20 \choose 17}=1140$, which is in turn much smaller than $N(\mA)$.

In general, for a fixed $k$, the difference $N(\mA)-N_{\max}(\mA)\rightarrow \infty$ as $N \rightarrow \infty$. Therefore, the improvement of $N_{\max}(\mA)$ over $N(\mA)$ is unbounded.
\end{eg}

\section{Efficient Algorithm for Computing the Lower Bound}\label{section_algorithms}

In Section~\ref{section_bounds}, a new lower bound on the required alphabet size of the existence of secure network codes over a wiretap network $(G, \mA)$ is obtained. This lower bound is graph-theoretical, and specifically it depends on the topology of the network $G$ and the collection $\mA$ of wiretap sets. However, it is not given in a form which is readily computable. In this section, we develop a polynomial-time algorithm to compute this lower bound.

\subsection{Primary Minimum Cut}

\begin{defn}[\textbf{Primary Minimum Cut}]\label{defn_primary-mincut}\em
Consider a finite directed acyclic network $G=(V, E)$ with a single source $s$, and let $t$ be a non-source node in $V$. A minimum cut between $s$ and $t$ in $G$ is primary, if it separates $s$ and all the minimum cuts between $s$ and $t$. In other words, a primary minimum cut between $s$ and $t$ is a common minimum cut of all the minimum cuts between $s$ and $t$.
\end{defn}

The notion of primary minimum cut is crucial to the development of our algorithm in the next subsection. We will first prove the existence and uniqueness of the primary minimum cut between the source node $s$ and a non-source node $t$. In the following, we introduce the binary relation ``$\leq$'' amongst the minimum cuts between $s$ and $t$.

\begin{defn}\label{defn_relation-cuts}\em
Let $\CUT_1$ and $\CUT_2$ be two minimum cuts between $s$ and $t$. We write $\CUT_1\leq \CUT_2$, if $\CUT_1$ separates $\CUT_2$ from $s$, or equivalently, $\CUT_1$ is a cut between $s$ and $\CUT_2$.
\end{defn}

This binary relation ``$\leq$'' between two minimum cuts is a non-strict partial order (Definition~\ref{defn_2_partial_orders}), as to be proved in the following theorem. This further implies that for two distinct minimum cuts $\CUT_1$ and $\CUT_2$ between $s$ and $t$, $\CUT_2 \nleq \CUT_1$ provided that $\CUT_1 \leq \CUT_2$.

\begin{thm}\label{lem_partial-order-CutRelation}\em
The binary relation ``$\leq$'' amongst the minimum cuts between $s$ and $t$ is a non-strict partial order.
\end{thm}
\begin{IEEEproof}
Denote by $\MinCut(t)$ the set of all minimum cuts between $s$ and $t$, and let $\CUT_1$, $\CUT_2$, and $\CUT_3$ be three minimum cuts in $\MinCut(t)$. Reflexivity is apparent. For antisymmetry, we assume $\CUT_1 \leq \CUT_2$ and $\CUT_2 \leq \CUT_1$. By Definition~\ref{defn_relation-cuts}, we obtain that $\CUT_1$ (resp. $\CUT_2$) separates $\CUT_2$ (resp. $\CUT_1$) from $s$. This implies $\CUT_1=\CUT_2$.

To prove transitivity, i.e., if $\CUT_1 \leq \CUT_2$ and $\CUT_2 \leq \CUT_3$, then $\CUT_1 \leq \CUT_3$, we discuss the following two cases:
\begin{description}
  \item[{\bf Case 1.}]\quad At least two out of the three minimum cuts are the same. Transitivity is immediate.
  \item[{\bf Case 2.}]\quad The three minimum cuts are distinct. Then, $\CUT_1\leq \CUT_2$ and $\CUT_2\leq \CUT_3$ mean that $\CUT_1$ separates $\CUT_2$ from $s$ and $\CUT_2$ separates $\CUT_3$ from $s$, respectively. Consequently, $\CUT_1$ separates $\CUT_3$ from $s$, which proves $\CUT_1 \leq \CUT_3$ by Definition~\ref{defn_relation-cuts}.
\end{description}
The theorem is proved.
\end{IEEEproof}


The proof of the proposition below is straightforward and so it is omitted.

\begin{prop}\label{prop}\em
Let $n=\mincut(s, t)$ and $\CUT$ be an arbitrary minimum cut between $s$ and $t$. Then an arbitrary set of $n$ edge-disjoint paths from $s$ to $t$ contains all the $n$ edges in $\CUT$, and each of the $n$ edges is on an exactly one of the $n$ edge-disjoint paths.
\end{prop}

For an acyclic network $G$, there exists an {\em upstream-to-downstream order} (also called {\em ancestral topological order}) on the edges in $E$, which is consistent with the natural partial order of the edges. To be specific, for two distinct edges $d$ and $e$ in $E$, if there is a directed path from $d$ to $e$, we write $d\leq e$.\footnote{Here we use the same symbol ``$\leq$'' to represent two binary relations, but this abuse of notation should cause no ambiguity.} We also set $e\leq e$, $\forall e\in E$. It is not difficult to see that the binary relation ``$\leq$'' amongst the edges in $E$ is a non-strict partial order: the reflexivity and transitivity of ``$\leq$'' are immediate, and the antisymmetry follows from the acyclicity of $G$. The following lemma provides a necessary and sufficiency condition for $\CUT_1 \leq \CUT_2$, where $\CUT_1$ and $\CUT_2$ are two minimum cuts between $s$ and $t$.

\begin{lemma}\label{prop1_relation-cuts}\em
Let $\CUT_1, \CUT_2 \in \MinCut(t)$, the set of all minimum cuts between $s$ and $t$. Then $\CUT_1\leq \CUT_2$ if and only if there exist $n=\mincut(s, t)$ edge-disjoint paths $P_1, P_2,\cdots,P_n$ from $s$ to $t$ such that $e_{1,i}\leq e_{2,i}$, where $P_i\cap \CUT_1=\{ e_{1,i} \}$ and $P_i\cap \CUT_2=\{ e_{2,i} \}$ for all $1 \leq i \leq n$.
\end{lemma}

Following Proposition~\ref{prop} and Lemma~\ref{prop1_relation-cuts}, the next theorem asserts that the order between two minimum cuts under the relation ``$\leq$'' is independent of which set of $n$ edge-disjoint paths from $s$ to $t$ is chosen. The proofs of Lemma~\ref{prop1_relation-cuts} and Theorem~\ref{thm_relation-cuts} are relegated to Appendix~\ref{app:c}.

\begin{thm}\label{thm_relation-cuts}\em\
Let $n=\mincut(s, t)$ and $\CUT_1, \CUT_2 \in \MinCut(t)$ with $\CUT_1\leq \CUT_2$, and $P_1, P_2,\cdots,P_n$ be $n$ arbitrary edge-disjoint paths from $s$ to $t$. Then $e_{1,i}\leq e_{2,i}$ holds for all $i=1, 2, \cdots, n$, where $P_i \cap \CUT_1=\{ e_{1,i} \}$ and $P_i \cap \CUT_2=\{ e_{2,i} \}$, $1 \leq i \leq n$.
\end{thm}

We now proceed to prove the existence and uniqueness of the primary minimum cut in Definition~\ref{defn_primary-mincut}.

\begin{enumerate}
  \item \textbf{Existence:} Let $\CUT_1=\{e_{1,i}: i=1,2, \ldots, n \}$ and $\CUT_2=\{e_{2,i}: i=1,2, \ldots, n  \}$ be two minimum cuts in $\MinCut(t)$, and $P_1, P_2, \cdots, P_n$ be $n$ arbitrary edge-disjoint paths from $s$ to $t$, where $n=\mincut(s,t)$. We assume without loss of generality that $P_i\cap \CUT_1=\{ e_{1,i} \}$ and $P_i\cap \CUT_2=\{ e_{2,i} \}$ for $1\leq i \leq n$. Define an edge set
$$\CUT=\{e_{i}=\minord(e_{1,i}, e_{2,i}):\ i=1,2, \ldots, n  \},$$
where
\begin{align*}
\minord(e_{1,i}, e_{2,i})=
\begin{cases} e_{1,i}, & \text{if } e_{1,i} \leq e_{2,i};\\
              e_{2,i}, & \text{otherwise.}
\end{cases}
\end{align*}
It was proved in \cite[Lemma 5]{Guang-SmlFieldSize-SNC-comm-lett} that the above edge set $\CUT$ is also a minimum cut between $s$ and $t$. By Lemma~\ref{prop1_relation-cuts}, we further have $\CUT\leq \CUT_1$ and $\CUT\leq \CUT_2$. Consequently, by Definition~\ref{defn_relation-cuts}, $\CUT$ is a common minimum cut of $\CUT_1$ and $\CUT_2$. Thus, we obtain $\CUT_1\sim \CUT_2$ by the definition of the equivalence relation ``$\sim$''. Similarly, we can prove that all the minimum cuts in $\MinCut(t)$ are equivalent. We also have $|\MinCut(t)|<\infty$ since the network $G$ is finite. Thus, by the argument immediately before Proposition~\ref{lem_1}, there exists a common minimum cut $\CUT^*$ of all minimum cuts in $\MinCut(t)$. In other words, $\CUT^*$ is a primary minimum cut between $s$ and $t$ by Definition~\ref{defn_primary-mincut}.
  \item \textbf{Uniqueness:} Let $\CUT_1^*$ and $\CUT_2^*$ be two primary minimum cuts between $s$ and $t$. We obtain that $\CUT_1^* \leq \CUT_2^*$ and $\CUT_2^* \leq \CUT_1^*$ by Definitions~\ref{defn_primary-mincut}~and~\ref{defn_relation-cuts}. This implies $\CUT_1^*=\CUT_2^*$ by the antisymmetry of ``$\leq$'' in Theorem \ref{lem_partial-order-CutRelation}.
  \end{enumerate}

The concept of the primary minimum cut between the source node $s$ and a non-source node $t$ can be extended to between $s$ and a wiretap set $A\in \mA$ in the same way that the concept of a cut between $s$ and $t$ is extended to between $s$ and $A$. In particular, for every wiretap set $A\in \mA$, there exists a unique primary minimum cut between $s$ and $A$, and further the minimum cut between $s$ and the primary minimum cut is unique, i.e., itself.

Based on the above discussions, we now prove the next theorem, which shows that the computation of $N_{\max}(\mA)$ can be reduced to the computation of a set of primary minimum cuts such that each wiretap set $A\in \mA$ ($A$ is assumed to be regular) is separated from $s$ by at least one primary minimum cut in this set. This theorem is the cornerstone in the development of our efficient algorithm to compute $N_{\max}(\mA)$.

\begin{thm}\label{thm_primary-mincut}\em
Let $A$ be a regular edge set in a finite directed acyclic network $G$ with a single source node $s$, and $\CUT$ be the primary minimum cut between $s$ and $A$. Then, the following hold:
\begin{enumerate}
  \item For any regular edge set $A'$ with $A'\sim A$, $\CUT$ is also the primary minimum cut between $s$ and~$A'$.\label{con_1}
  \item For any regular edge set $B$ with $B\prec A$, $\CUT$ separates $B$ from $s$.\label{con_2}
\end{enumerate}
\end{thm}
\begin{IEEEproof}
Let $A'$ be a regular edge set with $A' \sim A$ and $A' \neq A$, and $\CUT'$ be the primary minimum cut between $s$ and $A'$. We now prove that $\CUT=\CUT'$ as follows. First we see that $\CUT\sim A$ and $\CUT'\sim A'$. Together with $A\sim A'$, $\CUT\sim \CUT'$ follows from the transitivity of ``$\sim$''. Thus, $\CUT$ and $\CUT'$ have a common minimum cut, denoted by $\CUT^*$. While $\CUT^*$ separates $\CUT$ from $s$, $\CUT^*$ is also a minimum cut between $s$ and $A$. Then it follows from Definition \ref{defn_relation-cuts} that
\begin{align}\label{equ1_thm_primary-mincut}
\CUT^*\leq \CUT.
\end{align}
On the other hand, since $\CUT$ is the primary minimum cut between $s$ and $A$, and $\CUT^*$ is a minimum cut between $s$ and $A$, we also have $\CUT \leq \CUT^*$ by Definition \ref{defn_primary-mincut}. Combining this with \eqref{equ1_thm_primary-mincut}, we obtain $\CUT=\CUT^*$. Similarly, we can prove that $\CUT'=\CUT^*$. Therefore, $\CUT=\CUT'$.

Next, we prove \ref{con_2}). Since $B \prec A$, by Definition~\ref{defn_domination}, there exists a minimum cut $\CUT''$ of $A$ which separates $B$ from $s$. Furthermore, since $\CUT$ is the primary minimum cut between $s$ and $A$, we can see that $\CUT \leq \CUT''$ by Definitions~\ref{defn_primary-mincut}~and~\ref{defn_relation-cuts}. This implies that $\CUT$ separates $B$ from $s$. The theorem is proved.
\end{IEEEproof}

\begin{cor}\label{cor_primary-mincut}\em
In a wiretap network $(G, \mA)$, let $\Cl$ be an arbitrary equivalence class of the wiretap sets. Then
\begin{enumerate}
  \item all the wiretap sets in $\Cl$ have the same primary minimum cut, which hence is called the primary minimum cut of the equivalence class $\Cl$, and
  \item for every equivalence class $\Cl'$ with $\Cl' \prec \Cl$, the primary minimum cut of $\Cl$ separates all the wiretap sets in $\Cl'$ from $s$.
\end{enumerate}
\end{cor}

The above corollary can be proved by a straightforward application of Theorem~\ref{thm_primary-mincut}. Since two maximal equivalence classes in $(G, \mA)$ cannot share a common primary minimum cut, this corollary shows that in order to compute $N_{\max}(\mA)$ for a wiretap network $(G, \mA)$, it suffices to find the primary minimum cuts of all the maximal equivalence classes in $(G, \mA)$.

\subsection{Algorithm}

Based on the observation at the end of the last subsection, we now develop an efficient algorithm for computing $N_{\max}(\mA)$. In our algorithm, the primary minimum cuts of all the maximal equivalence classes in $(G, \mA)$ are obtained without first determining the equivalence classes of wiretap sets and the domination relation among them. This is the key to the efficiency of the algorithm. To be specific, we compute $N_{\max}(\mA)$ as follows:
\begin{enumerate}
  \item Define a set $\mB$, and initialize $\mB$ to the empty set.\label{step-1}
  \item Arbitrarily choose a wiretap set $A\in \mA$ that has the largest cardinality in $\mA$. Find the primary minimum cut between $s$ and $A$, and call it $\CUT$.\label{step-2}
  \item Partition the edge set $E$ into two disjoint subsets: $E_{\CUT}$ and $E_{\CUT}^c\triangleq E\setminus E_{\CUT}$, where $E_{\CUT}$ is the set of the edges reachable from the source node $s$ upon deleting the edges in $\CUT$. Note that $\CUT\subset E_{\CUT}^c$. \label{step-3}
  \item Remove all the wiretap sets in $\mA$ that are subsets of $E_{\CUT}^c$ and add the primary minimum cut $\CUT$ to $\mB$.\label{step-4}
  \item Repeat Steps \ref{step-2}) to \ref{step-4}) until $\mA$ is empty and output $\mB$, where $N_{\max}(\mA)=|\mB|$.\label{step-5}
\end{enumerate}

The algorithm is explained as follows:
\begin{itemize}
\item In Step \ref{step-2}), since the algorithm always chooses a wiretap set $A\in \mA$ that has the largest cardinality in $\mA$, the chosen wiretap set $A$ belongs to a maximal equivalence class.
\item In Step \ref{step-3}), according to Corollary~\ref{cor_primary-mincut}, the wiretap sets in the equivalence class $\Cl(A)$ or an equivalence class $\Cl$ with $\Cl\prec\Cl(A)$ are subsets of $E_{\CUT}^c$. Removing these wiretap sets from $\mA$ is equivalent to removing $\Cl(A)$ and all equivalence classes $\Cl$ with $\Cl\prec \Cl(A)$.
\item In addition, for any other equivalence class $\Cl'$ with $\Cl' \nprec \Cl(A)$, by Theorem~\ref{cor_1}, we have $A'\nprec A$ for any wiretap set $A'\in \Cl'$. We now prove by contradiction that $A'\nprec \CUT$. Assume that $A' \prec \CUT$. Then there exists a minimum cut $\CUT^*$ of $\CUT$ that separates $s$ from $A'$. Since $\CUT$ is the primary minimum cut of $A$ and $\CUT^*$ is a minimum cut of $\CUT$, we have $\CUT^*=\CUT$ and $\CUT^*$ separates $s$ from $A$. Then $\CUT^*$ is a (primary) minimum cut of $A$ that separates $s$ from $A'$, implying that $A'\prec A$, which is a contradiction to $A'\nprec A$.
\item As such, none of the wiretap sets in $\Cl' \nprec\Cl(A)$ are removed from $\mA$, and in particular, all maximal equivalence classes other than $\Cl(A)$ are not removed from $\mA$. Thus, exactly one maximal equivalence class is removed from $\mA$ in each iteration.
\end{itemize}

\begin{rem}\em
If in Step~\ref{step-4}) we instead consider only those wiretap sets of the same cardinality as $A$, which means that only the wiretap sets in $\Cl(A)$ are removed from $\mA$, then the algorithm at the end outputs $\mB$ with $|\mB|=N(\mA)$ instead of $N_{\max}(\mA)$.
\end{rem}

In the proposed algorithm, we assume that all the wiretap sets in $\mA$ are regular. The algorithm can be modified so that it continues to be applicable without this assumption. This can be done by replacing ``arbitrarily choose a wiretap set $A\in \mA$ that has the largest cardinality in $\mA$'' in Step \ref{step-2}) by ``arbitrarily choose a wiretap set $A\in \mA$ that has the largest minimum cut capacity in $\mA$''. However, this would require pre-computing the minimum cut capacity of every wiretap set in $\mA$ (this is essentially the same as replacing every non-regular wiretap set in $\mA$ by one of its minimum cuts, which is regular). Although the complexity for computing the minimum cut capacity of a wiretap set is only polynomial in $|E|$, this will still significantly increase the computational complexity of the algorithm when $|\mA|$ is large. To avoid this shortcoming, we modify the original algorithm (which assumes that all the wiretap sets are regular) by replacing Step \ref{step-4}) by:
\begin{description}
\item[4')] Remove all the wiretap or edge sets in $\mA\cup\mB$ that are subsets of $E_{\CUT}^c$. Add the primary minimum cut $\CUT$ to $\mB$.
\end{description}
An implementation of this algorithm can be found in Algorithm~\ref{algo_N-max-A-2}. Before we explain this algorithm, we first generalize two definitions and prove a lemma.

\begin{algorithm}[!htb]\label{algo_N-max-A-2}
\SetAlgoLined
\KwIn{The wiretap network $(G, \mA)$, where $G=(V, E)$.}
\KwOut{$N_{\max}(\mA)$, the number of maximal equivalence classes with respect to $(G, \mA)$.}
\BlankLine
\Begin{
\nl Set $\mB=\emptyset$\;
\nl \While{$\mA \neq \emptyset$}{
\nl      choose a wiretap set $A$ of the largest cardinality in $\mA$\;
\nl      find the primary minimum cut $\CUT$ of $A$\;
\nl      partition $E$ into two parts $E_{\CUT}$ and $E_{\CUT}^c=E\setminus E_{\CUT}$\;
\nl      \For{ each $B\in \mA\cup\mB$ }{
\nl      \If{ $B\subseteq E_{\CUT}^c$ }{
\nl      remove $B$ from $\mA$.}}
\nl      add $\CUT$ to $\mB$.}
\nl      Return $\mB$. \tcp*[f]{\rm Note that $|\mB|=N_{\max}(\mA)$.}}
\caption{Algorithm for Computing $N_{\max}(\mA)$}
\end{algorithm}

\begin{defn}\label{defn:A}\em
Two wiretap sets are equivalent if they have a common minimum cut.
\end{defn}

\begin{defn}\label{defn:B}\em
Let $A'$ and $A$ be two wiretap sets in $\mA$. Then $A'$ is dominated by $A$ (write $A'\prec A$) if $\mincut(s, A')<\mincut(s, A)$ and there exists a minimum cut between $s$ and $A$ that separates $A'$~from~$s$.
\end{defn}

Definitions~\ref{defn:A}~and~\ref{defn:B} are generalizations of their original versions that require the wiretap sets to be regular.

\begin{lemma}\label{lem_3}\em
  For two wiretap sets $A_1$ and $A_2$ (not necessarily regular) in $\mA$ with $\mincut(s, A_1)<\mincut(s, A_2)$, $A_1\prec A_2$ if and only if $\CUT_{A_1} \prec \CUT_{A_2}$, where $\CUT_{A_1}$ and $\CUT_{A_2}$ are the primary minimum cuts of $A_1$ and $A_2$, respectively.
  \end{lemma}
\begin{IEEEproof}
    For the ``if'' part, since $\CUT_{A_2}$ is the primary minimum cut of $A_2$ and $\CUT_{A_1} \prec \CUT_{A_2}$, $\CUT_{A_2}$ separates $\CUT_{A_1}$ from $s$. Together with $\CUT_{A_1}$ being the (primary) minimum cut of $A_1$, $\CUT_{A_2}$ separates $A_1$ from $s$, implying $A_1\prec A_2$. For the ``only if'' part, since $A_1\prec A_2$, there exists a minimum cut $\CUT_{A_2}^*$ between $s$ and $A_2$ that separates $A_1$ from $s$. This further implies that $\CUT_{A_2}^*$ is also a minimum cut between $s$ and $A_1\cup A_2$. By Lemma~\ref{lem_2}, we have $\CUT_{A_1} \prec \CUT_{A_2}^*$. In addition, since $\CUT_{A_2}$ is the primary minimum cut of $A_2$, we obtain $\CUT_{A_2}\leq \CUT_{A_2}^*$ by Definitions~\ref{defn_primary-mincut}~and~\ref{defn_relation-cuts}. Combining $\CUT_{A_1}\prec \CUT_{A_2}^*$ and $\CUT_{A_2}\leq \CUT_{A_2}^*$, we have proved that $\CUT_{A_1}\prec \CUT_{A_2}$.
\end{IEEEproof}

We now explain Algorithm~\ref{algo_N-max-A-2} as follows:

\begin{itemize}
  \item In Step 4'), $\CUT_{A}$ is added to $\mB$.
  \item If $A$ has the largest minimum cut capacity in $\mA$, then $\CUT_{A}$ will stay in $\mB$ until the algorithm terminates. This implies that $A$ belongs to a maximal equivalent class. Otherwise, there must exist a wiretap set $A'$ in $\mA$ such that $A'\succ A$. By Definition~\ref{defn:B}, this implies that $\mincut(s, A')>\mincut(s, A)$, contradicting the fact that $A$ has the largest minimum cut capacity in $\mA$.
  \item If $A$ does not have the largest minimum cut capacity in $\mA$,
    \begin{enumerate}
      \item if $A$ belongs to a maximal equivalence class (e.g. $\Cl_{11}$ in Example~\ref{eg_Ordered-Set-2}), then by Lemma~\ref{lem_3}, $\CUT_{A}$ will stay in $\mB$ until the algorithm terminates;
      \item otherwise (e.g. $\Cl_{12}$ in Example~\ref{eg_Ordered-Set-2}), by Lemma~\ref{lem_3}, $\CUT_A$ will subsequently be replaced by some primary minimum cut $\CUT_{A''}$ of a wiretap set $A''$ in $\mA$. Repeat this argument if necessary until a primary minimum cut $\CUT_{A^*}$ is added to $\mB$, where $\Cl(A^*)$ is a maximal equivalence class such that $A^* \succ A$.
    \end{enumerate}
  \item Combining all the above, we see that at the end the algorithm outputs $\mB$ that contains all the primary minimum cuts of the maximal equivalence classes in $\mA$, and computes the minimum cut capacity of a wiretap set for at most $N(\mA)$ times (instead of $|\mA|$ times).
\end{itemize}

In Algorithm~\ref{algo_N-max-A-2}, two key steps, namely finding the primary minimum cut and the edge partition (Lines 4 and 5 in Algorithm~\ref{algo_N-max-A-2}, respectively), are involved. The edge partition can be implemented efficiently by slightly modifying existing search algorithms on directed graphs \cite{Book-NetwFlow, Book-GraphTh-Bondy-Murty}. We can use a classical search algorithm to find all the nodes reachable along directed paths from the source node $s$. To find all the edges in $E_{\CUT}$, i.e., the edges reachable from $s$ upon deleting the edges in $\CUT$, it suffices to add a simple functionality for storing the reachable edges during the search process. The implementation is given in Algorithm~\ref{algo_search}. In \cite{Book-NetwFlow}, it is shown that the search algorithm runs in $\mO(|E|)$ time because in the worst case the algorithm needs to traverse all the edges in $E$. Here, since the primary minimum cut $\CUT$ of the wiretap set $A$ is removed from the network $G$, Algorithm~\ref{algo_search} can find the edge set $E_{\CUT}$ in $\mO(|E_{\CUT}|)$ time.

\begin{algorithm}[!htb]\label{algo_search}
\SetAlgoLined
\Begin{
\nl Unmark all nodes in $V$\;
\nl mark source node $s$\;
\nl $\pred(s):=0$\tcp*[r]{\rm $\pred(i)$ refers to a predecessor node of node $i$.}
\nl set the edge-set $\SET=\emptyset$\;
\nl set the node-set $\LIST=\{s\}$\;
\nl \While{$\LIST\neq \emptyset$}{
\nl      select a node $i$ in $\LIST$\;
\nl      \eIf{ node $i$ is incident to an edge $(i,j)$ such that node $j$ is unmarked}{
\nl          mark node $j$\;
\nl          $\pred(j):=i$\;
\nl          add node $j$ to $\LIST$\;
\nl          add all parallel edges leading from $i$ to $j$ to $\SET$\;
}{
\nl          delete node $i$ from $\LIST$\;
}
         }
\nl Return the edge-set $\SET$.
         }
\caption{Search Algorithm}
\end{algorithm}

Before giving an efficient algorithm for finding the primary minimum cut, we first introduce some notation below. Let $G=(V, E)$ be a directed acyclic network with a single source node $s$ and $t$ be a sink node in $V\setminus \{s\}$. Denote by $C_t$ the minimum cut capacity between the source node $s$ and the sink node~$t$. By the max-flow min-cut theorem \cite{maxflow-Ford-Fulkerson,Elias-Feinstein-Shannon-maxflow}, the value $v(f)$ of a maximum flow $f$ from $s$ to $t$ is equal to the minimum cut capacity $C_t$ between $s$ and $t$, i.e., $v(f)=C_t$. Since all the edges in the network $G$ have unit-capacity (i.e., the capacity is $1$), $C_t$ is a positive integer and the maximum flow $f$ can be decomposed into $C_t$ edge-disjoint paths from $s$ to $t$. Various algorithms for finding such edge-disjoint paths can be implemented in polynomial time in $|E|$ \cite{Book-NetwFlow, Book-GraphTh-Bondy-Murty}.

Now, we explore efficient algorithms for finding the primary minimum cut between $s$ and an edge set. For the convenience of presentation, we instead consider algorithms for finding the primary minimum cut between $s$ and a node $t\neq s$. Toward this end, we propose Algorithm~\ref{algo_Augmenting-Path} which takes as input a set of $C_t$ edge-disjoint paths from $s$ to $t$. Such a set of paths can be obtained by using any existing algorithm for this purpose. The verification of Algorithm~\ref{algo_Augmenting-Path} is given in Appendix~\ref{app:d}. We give an example below to illustrate the algorithm.

\begin{algorithm}[!htb]\label{algo_Augmenting-Path}
\SetAlgoLined
\KwIn{An acyclic network $G=(V, E)$ with a maximal flow $f$ from the source node $s$ to a sink node $t$, i.e., for every edge $e$ in the corresponding $C_t$ edge-disjoint paths, the flow value is defined as $1$, written as $f(e)=1$; otherwise, the flow value is defined as $0$, written as $f(e)=0$.}
\KwOut{The primary minimum cut between $s$ and $t$.}
\BlankLine
\Begin{
\nl Set $S=\{ s \}$\;
\nl \For{ each node $i\in S$ }{
\nl    \If{ $\exists$ a node $j\in V \setminus S$ s.t. either $\exists$ a forward edge $e$ from $i$ to $j$ s.t. $f(e)=0$ or $\exists$ a reverse edge $e$ from $j$ to $i$ s.t. $f(e)=1$ 
}{
\nl      replace $S$ by $S \cup \{j\}$.}}
\nl Return $\CUT=\{e:~\tail(e)\in S \text{ and } \head(e)\in V\setminus S\}$.
       }
\caption{Algorithm for Finding the Primary Minimum Cut}
\end{algorithm}

\begin{eg}\em

\begin{figure}[!h]
\centering
\subfigure[Initialize $S$ to $\{s\}$.]{
 \label{subfig:primary_mincut_a}
\begin{tikzpicture}
[->,>=stealth',shorten >=1pt,auto,node distance=1.6cm]
  \tikzstyle{every state}=[fill=none,draw=black,text=black,minimum size=6pt,inner sep=0pt]
  \tikzstyle{inode}=[draw,circle,fill=black,minimum size=6pt, inner sep=0pt]
  \node[inode]         (s)[label=right:$s$]                 {};
  \node[state]         (i_1)[below left of=s, xshift=-7mm, label=left:$i_1$]   {};
  \node[state]         (i_2)[below right of=i_1, label=right:$i_2$] {};
  \node[state]         (i_3)[right of=i_2, xshift=5mm, label=right:$i_3$] {};
  \node[state]         (i_4)[right of=i_3, label=right:$i_4$] {};
  \node[state]         (i_6)[below of=i_2, label=right:$i_6$] {};
  \node[state]         (i_5)[below left of=i_6, label=left:$i_5$] {};
  \node[state]         (i_7)[below of=i_3, label=right:$i_7$] {};
  \node[state]         (i_8)[below of=i_4, label=right:$i_8$] {};
  \node[state]         (i_9)[below of=i_6, label=right:$i_9$] {};
  \node[state]         (i_{10})[below of=i_7, label=left:$i_{10}$] {};
  \node[state]         (i_{11})[below of=i_8, label=right:$i_{11}$] {};
  \node[state]         (t)[below of=i_9, label=left:$t$] {};
\path
(s) edge           node[pos=0.4, left=0.5mm] {$1$} (i_1)
    edge           node[pos=0.5, left=-1mm] {$1$} (i_2)
    edge           node[pos=0.5, right=-0.5mm] {$0$} (i_3)
    edge           node[pos=0.5, right=0mm] {$1$} (i_4)
    edge           node[pos=0.5, left=-1mm] {$1$} (i_7)
(i_1) edge       node[pos=0.6, left=0mm] {$0$}(i_2)
      edge       node[pos=0.5, left=-1mm]{$1$}(i_5)
(i_2) edge       node[pos=0.5, left=-1mm]{$1$}(i_6)
(i_3) edge       node[pos=0.5, right=-1mm]{$0$}(i_7)
      edge       node[pos=0.5, right=-0.5mm]{$0$}(i_9)
(i_4) edge       node[pos=0.5, right=-1mm]{$1$}(i_8)
      edge       node[pos=0.3, left=-0.5mm]{$0$}(i_{10})
(i_5) edge       node[pos=0.5, left=-0.5mm]{$1$}(t)
(i_6) edge       node[pos=0.3, left=-0.5mm]{$0$}(i_5)
      edge       node[pos=0.5, left=-1mm]{$1$}(i_9)
(i_7) edge       node[pos=0.5, left=-1mm]{$1$}(i_{10})
(i_8) edge       node[pos=0.4, right=-0.5mm]{$0$}(i_{10})
      edge       node[pos=0.5, right=-1mm]{$1$}(i_{11})
(i_9) edge       node[pos=0.4, left=-1mm]{$1$}(t)
(i_{10}) edge    node[pos=0.4, left=-0.3mm]{$1$}(t)
(i_{11}) edge    node[pos=0.5, right=2mm]{$1$}(t);
\end{tikzpicture}
}
\hspace{1in}
\subfigure[Update $S$ to $\{ s, i_3, i_7, i_9 \}$.]{
\label{subfig:primary_mincut_b}
\begin{tikzpicture}
[->,>=stealth',shorten >=1pt,auto,node distance=1.6cm]
  \tikzstyle{every state}=[fill=none,draw=black,text=black,minimum size=6pt,inner sep=0pt]
  \tikzstyle{inode}=[draw,circle,fill=black,minimum size=6pt, inner sep=0pt]
  \node[inode]         (s)[label=right:$s$]                 {};
  \node[state]         (i_1)[below left of=s, xshift=-7mm, label=left:$i_1$]   {};
  \node[state]         (i_2)[below right of=i_1, label=right:$i_2$] {};
  \node[inode]         (i_3)[right of=i_2, xshift=5mm, label=right:$i_3$] {};
  \node[state]         (i_4)[right of=i_3, label=right:$i_4$] {};
  \node[state]         (i_6)[below of=i_2, label=right:$i_6$] {};
  \node[state]         (i_5)[below left of=i_6, label=left:$i_5$] {};
  \node[inode]         (i_7)[below of=i_3, label=right:$i_7$] {};
  \node[state]         (i_8)[below of=i_4, label=right:$i_8$] {};
  \node[inode]         (i_9)[below of=i_6, label=right:$i_9$] {};
  \node[state]         (i_{10})[below of=i_7, label=left:$i_{10}$] {};
  \node[state]         (i_{11})[below of=i_8, label=right:$i_{11}$] {};
  \node[state]         (t)[below of=i_9, label=left:$t$] {};
\path
(s) edge           node[pos=0.4, left=0.5mm] {$1$} (i_1)
    edge           node[pos=0.5, left=-1mm] {$1$} (i_2)
    edge[ultra thick]           node[pos=0.5, right=-0.5mm] {$0$} (i_3)
    edge           node[pos=0.5, right=0mm] {$1$} (i_4)
    edge           node[pos=0.5, left=-1mm] {$1$} (i_7)
(i_1) edge       node[pos=0.6, left=0mm] {$0$}(i_2)
      edge       node[pos=0.5, left=-1mm]{$1$}(i_5)
(i_2) edge       node[pos=0.5, left=-1mm]{$1$}(i_6)
(i_3) edge[ultra thick]       node[pos=0.5, right=-1mm]{$0$}(i_7)
      edge[ultra thick]       node[pos=0.5, right=-0.5mm]{$0$}(i_9)
(i_4) edge       node[pos=0.5, right=-1mm]{$1$}(i_8)
      edge       node[pos=0.3, left=-0.5mm]{$0$}(i_{10})
(i_5) edge       node[pos=0.5, left=-0.5mm]{$1$}(t)
(i_6) edge       node[pos=0.3, left=-0.5mm]{$0$}(i_5)
      edge       node[pos=0.5, left=-1mm]{$1$}(i_9)
(i_7) edge       node[pos=0.5, left=-1mm]{$1$}(i_{10})
(i_8) edge       node[pos=0.4, right=-0.5mm]{$0$}(i_{10})
      edge       node[pos=0.5, right=-1mm]{$1$}(i_{11})
(i_9) edge       node[pos=0.4, left=-1mm]{$1$}(t)
(i_{10}) edge    node[pos=0.4, left=-0.3mm]{$1$}(t)
(i_{11}) edge    node[pos=0.5, right=2mm]{$1$}(t);
\end{tikzpicture}
}
\vspace{1in}
\subfigure[Update $S$ to $\{ s, i_2, i_3, i_5, i_6, i_7, i_9 \}$.]{
\label{subfig:primary_mincut_c}
\begin{tikzpicture}
[->,>=stealth',shorten >=1pt,auto,node distance=1.6cm]
  \tikzstyle{every state}=[fill=none,draw=black,text=black,minimum size=6pt,inner sep=0pt]
  \tikzstyle{inode}=[draw,circle,fill=black,minimum size=6pt, inner sep=0pt]
  \node[inode]         (s)[label=right:$s$]                 {};
  \node[state]         (i_1)[below left of=s, xshift=-7mm, label=left:$i_1$]   {};
  \node[inode]         (i_2)[below right of=i_1, label=right:$i_2$] {};
  \node[inode]         (i_3)[right of=i_2, xshift=5mm, label=right:$i_3$] {};
  \node[state]         (i_4)[right of=i_3, label=right:$i_4$] {};
  \node[inode]         (i_6)[below of=i_2, label=right:$i_6$] {};
  \node[inode]         (i_5)[below left of=i_6, label=left:$i_5$] {};
  \node[inode]         (i_7)[below of=i_3, label=right:$i_7$] {};
  \node[state]         (i_8)[below of=i_4, label=right:$i_8$] {};
  \node[inode]         (i_9)[below of=i_6, label=right:$i_9$] {};
  \node[state]         (i_{10})[below of=i_7, label=left:$i_{10}$] {};
  \node[state]         (i_{11})[below of=i_8, label=right:$i_{11}$] {};
  \node[state]         (t)[below of=i_9, label=left:$t$] {};
\path
(s) edge           node[pos=0.4, left=0.5mm] {$1$} (i_1)
    edge           node[pos=0.5, left=-1mm] {$1$} (i_2)
    edge[ultra thick]           node[pos=0.5, right=-0.5mm] {$0$} (i_3)
    edge           node[pos=0.5, right=0mm] {$1$} (i_4)
    edge           node[pos=0.5, left=-1mm] {$1$} (i_7)
(i_1) edge       node[pos=0.6, left=0mm] {$0$}(i_2)
      edge       node[pos=0.5, left=-1mm]{$1$}(i_5)
(i_2) edge[ultra thick]       node[pos=0.5, left=-1mm]{$1$}(i_6)
(i_3) edge[ultra thick]       node[pos=0.5, right=-1mm]{$0$}(i_7)
      edge[ultra thick]       node[pos=0.5, right=-0.5mm]{$0$}(i_9)
(i_4) edge       node[pos=0.5, right=-1mm]{$1$}(i_8)
      edge       node[pos=0.3, left=-0.5mm]{$0$}(i_{10})
(i_5) edge       node[pos=0.5, left=-0.5mm]{$1$}(t)
(i_6) edge[ultra thick]       node[pos=0.3, left=-0.5mm]{$0$}(i_5)
      edge[ultra thick]       node[pos=0.5, left=-1mm]{$1$}(i_9)
(i_7) edge       node[pos=0.5, left=-1mm]{$1$}(i_{10})
(i_8) edge       node[pos=0.4, right=-0.5mm]{$0$}(i_{10})
      edge       node[pos=0.5, right=-1mm]{$1$}(i_{11})
(i_9) edge       node[pos=0.4, left=-1mm]{$1$}(t)
(i_{10}) edge    node[pos=0.4, left=-0.3mm]{$1$}(t)
(i_{11}) edge    node[pos=0.5, right=2mm]{$1$}(t);
\end{tikzpicture}
}
\hspace{1in}
\subfigure[Update $S$ to $\{ s, i_1, i_2, i_3, i_5, i_6, i_7, i_9\}$.]{
\label{subfig:primary_mincut_d}
\begin{tikzpicture}
[->,>=stealth',shorten >=1pt,auto,node distance=1.6cm]
  \tikzstyle{every state}=[fill=none,draw=black,text=black,minimum size=6pt,inner sep=0pt]
  \tikzstyle{inode}=[draw,circle,fill=black,minimum size=6pt, inner sep=0pt]
  \node[inode]         (s)[label=right:$s$]                 {};
  \node[inode]         (i_1)[below left of=s, xshift=-7mm, label=left:$i_1$]   {};
  \node[inode]         (i_2)[below right of=i_1, label=right:$i_2$] {};
  \node[inode]         (i_3)[right of=i_2, xshift=5mm, label=right:$i_3$] {};
  \node[state]         (i_4)[right of=i_3, label=right:$i_4$] {};
  \node[inode]         (i_6)[below of=i_2, label=right:$i_6$] {};
  \node[inode]         (i_5)[below left of=i_6, label=left:$i_5$] {};
  \node[inode]         (i_7)[below of=i_3, label=right:$i_7$] {};
  \node[state]         (i_8)[below of=i_4, label=right:$i_8$] {};
  \node[inode]         (i_9)[below of=i_6, label=right:$i_9$] {};
  \node[state]         (i_{10})[below of=i_7, label=left:$i_{10}$] {};
  \node[state]         (i_{11})[below of=i_8, label=right:$i_{11}$] {};
  \node[state]         (t)[below of=i_9, label=left:$t$] {};
\path
(s) edge           node[pos=0.4, left=0.5mm] {$1$} (i_1)
    edge           node[pos=0.5, left=-1mm] {$1$} (i_2)
    edge[ultra thick]           node[pos=0.5, right=-0.5mm] {$0$} (i_3)
    edge           node[pos=0.5, right=0mm] {$1$} (i_4)
    edge           node[pos=0.5, left=-1mm] {$1$} (i_7)
(i_1) edge       node[pos=0.6, left=0mm] {$0$}(i_2)
      edge[ultra thick]       node[pos=0.5, left=-1mm]{$1$}(i_5)
(i_2) edge[ultra thick]       node[pos=0.5, left=-1mm]{$1$}(i_6)
(i_3) edge[ultra thick]       node[pos=0.5, right=-1mm]{$0$}(i_7)
      edge[ultra thick]       node[pos=0.5, right=-0.5mm]{$0$}(i_9)
(i_4) edge       node[pos=0.5, right=-1mm]{$1$}(i_8)
      edge       node[pos=0.3, left=-0.5mm]{$0$}(i_{10})
(i_5) edge       node[pos=0.5, left=-0.5mm]{$1$}(t)
(i_6) edge[ultra thick]       node[pos=0.3, left=-0.5mm]{$0$}(i_5)
      edge[ultra thick]       node[pos=0.5, left=-1mm]{$1$}(i_9)
(i_7) edge       node[pos=0.5, left=-1mm]{$1$}(i_{10})
(i_8) edge       node[pos=0.4, right=-0.5mm]{$0$}(i_{10})
      edge       node[pos=0.5, right=-1mm]{$1$}(i_{11})
(i_9) edge       node[pos=0.4, left=-1mm]{$1$}(t)
(i_{10}) edge    node[pos=0.4, left=-0.3mm]{$1$}(t)
(i_{11}) edge    node[pos=0.5, right=2mm]{$1$}(t);
\end{tikzpicture}
}
\vspace{-2cm}
\caption{An example to use Algorithm~\ref{algo_Augmenting-Path} for finding the primary minimum cut between $s$ and $t$ on the network $G$.}
\label{fig:primary_mincut}
\end{figure}
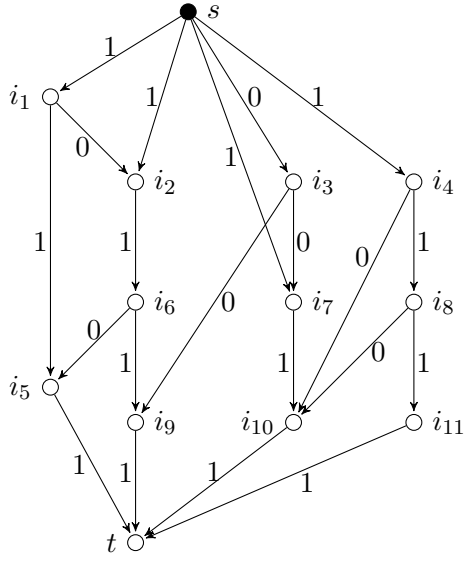
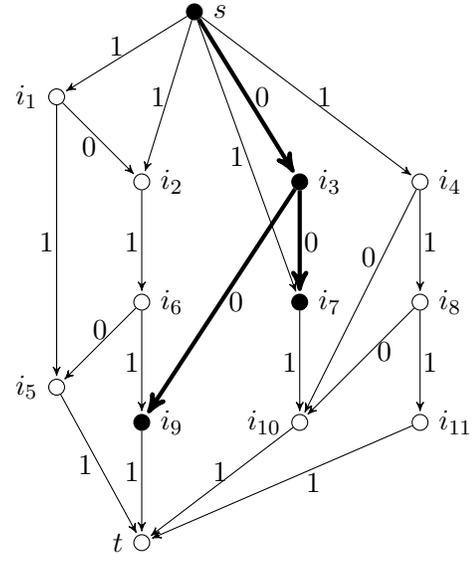
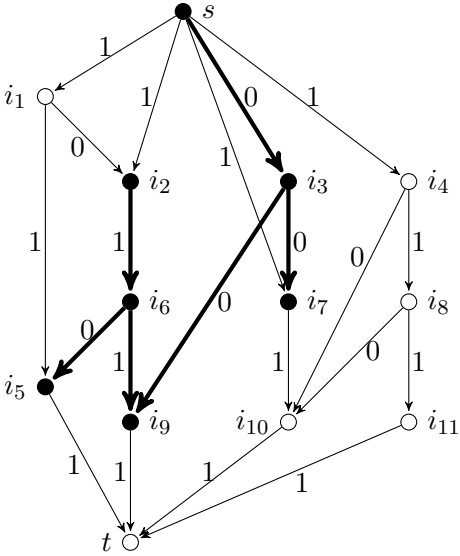
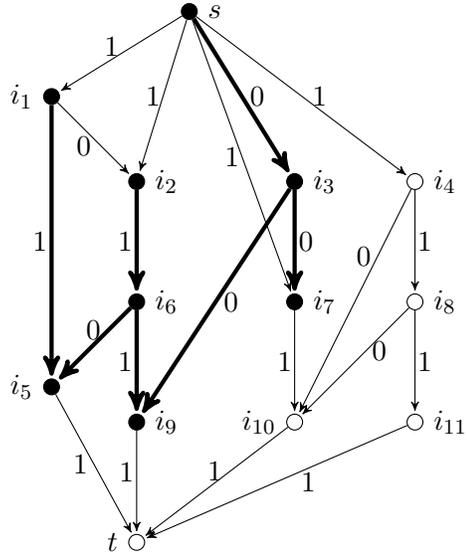

A directed acyclic network $G$ with a maximum flow $f$ from $s$ to $t$ is depicted in Fig.~\ref{subfig:primary_mincut_a}, where $s$ and $t$ are the source node and the sink node, respectively. Fig.~\ref{fig:primary_mincut} illustrates Algorithm~\ref{algo_Augmenting-Path} that outputs the primary minimum cut between $s$ and $t$ in $G$:
\begin{itemize}
  \item At first, only the pair of nodes $(s, i_3)$ satisfies Line 3 in Algorithm~\ref{algo_Augmenting-Path}, i.e., $\exists$ a forward edge from $s$ to $i_3$ with the flow value $0$. Update $S$ to $\{ s, i_3 \}$. Next, for $i_3$, $(i_3, i_7)$ and $(i_3, i_9)$ are two pairs of nodes such that there exist two forward edges with the flow value $0$ from $i_3$ to $i_7$ and $i_9$, respectively. Update $S$ to $\{ s, i_3, i_7, i_9 \}$. This is illustrated in Fig.~\ref{subfig:primary_mincut_b}.
  \item For $i_7\in S$, the edge $(i_7,i_{10})$ is the only edge connecting $i_7$ with another node in $V\setminus S$. It is a forward edge but with the flow value $1$ and does not satisfy Line 3 in Algorithm~\ref{algo_Augmenting-Path}. For $i_9$, since $i_9\in S$, $i_6\notin S$ and $(i_9, i_6)$ is a reverse edge with the flow value $1$, update $S$ to $\{ s, i_3, i_6, i_7, i_9 \}$. Similarly, we further update $S$ to $\{ s, i_2, i_3, i_5, i_6, i_7, i_9 \}$ by considering $i_6\in S$. This is illustrated in Fig.~\ref{subfig:primary_mincut_c}.
  \item Finally, since $(i_5, i_1)$ is a reverse edge with the flow value $1$, we obtain $S=\{ s, i_1, i_2, i_3, i_5, i_6, i_7, i_9\}$, as illustrated in Fig.~\ref{subfig:primary_mincut_d}.
\end{itemize}
Then the output edge set that is the primary minimum cut between $s$ and $t$ on $G$ is
\begin{align*}
\CUT=\{e:~\tail(e)\in S \text{ and } \head(e)\in V\setminus S\}=\{ (i_5, t), (i_9, t), (i_7, i_{10}), (s, i_4)\}.
 \end{align*}
\end{eg}

In fact, Algorithm~\ref{algo_Augmenting-Path} can be regarded as the last part of the augmenting path algorithm \cite{maxflow-Ford-Fulkerson,Elias-Feinstein-Shannon-maxflow} (also see \cite[Chapter 6.5]{Book-NetwFlow} and \cite[Chapter 7.2]{Book-GraphTh-Bondy-Murty}) for determining the termination of the algorithm, i.e., the flow value cannot be further increased. Algorithm~\ref{algo_Augmenting-Path} requires at most $\mO(|E|)$ time since the search method examines each edge at most once. If we use the augmenting path algorithm to find $C_t$ edge-disjoint paths from $s$ to $t$, then Algorithm~\ref{algo_Augmenting-Path} is already incorporated, and the total computational complexity for finding the primary minimum cut between $s$ and $t$ is at most $\mO(C_t\cdot|E|)$ since the path augmentation approach requires at most $\mO(|E|)$ time and the number of the path augmentations is upper bounded by the minimum cut capacity $C_t$. This total computational complexity may be reduced by employing more efficient maximum-flow algorithms for finding $C_t$ edge-disjoint paths from $s$ to $t$. For instance, if we suitably combine the features of the augmenting path algorithms and the shortest augmenting path algorithms \cite{Dinic-1970, Ahuja-Orlin-1991}, the total computational complexity is $\mO(\min\{ C_t^{2/3}|E|,\ |E|^{3/2}\})$, which is better than $\mO(C_t\cdot|E|)$.

Next, we continue to use the setup in Examples~\ref{eg_Ordered-Set-1} and~\ref{eg_Ordered-Set-2} to illustrate Algorithm~\ref{algo_N-max-A-2} for computing $N_{\max}(\mA)$.

\begin{eg}\em
Recall the wiretap network $(G,\mA)$ in Examples~\ref{eg_Ordered-Set-1} and \ref{eg_Ordered-Set-2}. Define a set $\mB$ and initialize $\mB$ to the empty set.
\begin{description}
  \item[Step 1:]\ \  Arbitrarily choose a wiretap set $A_1$ in $\mA$ of the largest cardinality $3$, for instance, $A_1=\{e_1, e_{11}, e_{16} \}$. Find $\CUT_{A_1}=\{ e_1, e_2, e_3 \}$, the primary minimum cut between $s$ and $A_1$, by Algorithm~\ref{algo_Augmenting-Path}. By Algorithm~\ref{algo_search}, we obtain the edge set \begin{align*}
      E_{\CUT_{A_1}}^c=\big\{ e_{1}, e_{2}, e_{3}, e_{6}, e_{7}, e_{8}, e_{9}, e_{10}, e_{11}, e_{16}, e_{18}, e_{19}\big\}.
      \end{align*}
      Remove the wiretap sets  from $\mA$ that are subsets of $E_{\CUT_{A_1}}^c$, e.g., $\{e_6\}$, $\{e_8, e_{18}\}$, $\{e_1, e_{11}, e_{16}\}$, etc. Update $\mA$ to
      \begin{align*}
      &\Big\{ \{ e_{12} \}, \{ e_{13} \}, \{ e_{14} \}, \{ e_{15} \}, \{ e_{20} \},\{ e_{21} \}, \{ e_{10}, e_{14}\}, \{ e_{10}, e_{15}\}, \{ e_{10}, e_{21}\},\\
          &\{ e_{11}, e_{14}\}, \{ e_{11}, e_{15}\}, \{ e_{11}, e_{20}\},
           \{ e_{12}, e_{20}\}, \{ e_{12}, e_{21}\}, \{ e_{13}, e_{17}\}, \{ e_{13}, e_{21}\},\\
            &\{ e_{14}, e_{20}\}, \{ e_{14}, e_{21}\}, \{ e_{15}, e_{20}\}, \{ e_{15}, e_{21}\}, \{ e_{18}, e_{20}\}, \{ e_{18}, e_{21}\}, \{ e_{19}, e_{20}\},\\
            &\{ e_{19}, e_{21}\}, \{ e_{3}, e_{5}, e_{17} \}, \{ e_{4}, e_{10}, e_{17} \}, \{ e_{5}, e_{11}, e_{17} \} \Big\}.
\end{align*}
We remark that every wiretap set in the updated $\mA$ has at least one edge not in $E_{\CUT_{A_1}}^c$. Add $\CUT_{A_1}$ to $\mB$ so that $\mB$ becomes $\{ \CUT_{A_1} \}$;
  \item[Step 2:] \ \  Arbitrarily choose a wiretap set $A_2$ in the updated $\mA$ of the largest cardinality $3$, say $A_2=\{e_4, e_{10}, e_{17} \}$, and then find its primary minimum cut $\CUT_{A_2}=\{ e_3, e_4, e_5 \}$ by Algorithm~\ref{algo_Augmenting-Path}. Then use Algorithm~\ref{algo_search} to obtain the edge set
      \begin{align*}
      E_{\CUT_{A_2}}^c=\big\{ e_{3}, e_{4}, e_{5}, e_{10}, e_{11}, e_{12}, e_{13}, e_{14}, e_{15}, e_{17}, e_{20}, e_{21} \big\},
      \end{align*}
      and remove the wiretap sets from $\mA$ that are subsets of $E_{\CUT_{A_2}}^c$. Update $\mA$ to
      \begin{align*}
      &\Big\{ \{ e_{18}, e_{20}\}, \{ e_{18}, e_{21}\}, \{ e_{19}, e_{20}\}, \{ e_{19}, e_{21}\} \Big\},
      \end{align*}
      and add $\CUT_{A_2}$ to $\mB$ so that $\mB$ becomes $\{ \CUT_{A_1}, \CUT_{A_2} \}$;
  \item[Step 3:]\ \ Arbitrarily choose a wiretap set $A_3=\{e_{19}, e_{20} \}$ in the updated $\mA$ of the largest cardinality $2$ and find its primary minimum cut $\CUT_{A_3}=\{ e_{16}, e_{17} \}$. All the wiretap sets in $\mA$ are subsets of $E_{\CUT_{A_3}}^c=\{ e_{16}, e_{17}, e_{18}, e_{19}, e_{20}, e_{21}\}$. Then update $\mA$ to the empty set and add $\CUT_{A_3}$ to $\mB$ so that $\mB$ becomes $\{ \CUT_{A_1}, \CUT_{A_2}, \CUT_{A_3} \}$.
\end{description}
Algorithm~\ref{algo_N-max-A-2} terminates and outputs $\mB$. Then we have $N_{\max}(\mA)=|\mB|=3$.
\end{eg}

\section{Conclusions}\label{section_conclusion}

In this paper, we have proved a new lower bound on the required alphabet size for the existence of secure network codes. Our lower bound depends only on the network topology and the collection of the wiretap sets. Our result shows that in general the required alphabet size can be reduced significantly without sacrificing security and information rate. Since our bound is not in closed form, we also have proposed a polynomial-time algorithm to compute it efficiently.

Toward developing our lower bound and the efficient algorithm for computing this bound, we have introduced/discussed various graph-theoretic concepts, including the equivalence relation between two edge sets (first appeared in \cite{Guang-SmlFieldSize-SNC-comm-lett}), the domination relation among equivalence classes of edge sets, the primary minimum cut between the source node and a sink node, etc. Although in this paper these concepts are applied solely in the context of secure network coding, they appear to be of fundamental interest in graph theory and we expect that they will find applications in graph theory and beyond.


\numberwithin{thm}{section}
\appendices

\section{Proof of Theorem~\ref{thm_cl-domination}}\label{app:a}

First assume $\Cl(A_1)\prec \Cl(A_2)$. By Definition~\ref{defn_cl-domin}, there exists a common minimum cut $\CUT$ of the wiretap sets in $\Cl(A_2)$ that also separates each of the wiretap sets in $\Cl(A_1)$ from $s$. For any $A_1'\in \Cl(A_1)$ and $A_2'\in \Cl(A_2)$, we have
\begin{align*}
|A_2|=\mincut(s, A_2')\leq \mincut(s, A_1'\cup A_2') \leq |\CUT|=|A_2|,
\end{align*}
where the first and the last equalities follows from $A_2 \sim A_2'$ and $A_2 \sim \CUT$, respectively. Together with Proposition~\ref{prop_wiretap-set}, this proves the ``only if'' part.

We next prove the ``if'' part. Assume that $A_1'\prec A_2'$ for any $A_1'\in\Cl(A_1)$ and $A_2'\in\Cl(A_2)$. For any $A_1'\in\Cl(A_1)$, since a minimum cut between $s$ and $A_1'\cup A_2$ is a cut between $s$ and $A_2$, together with the condition $A_1'\prec A_2$ and Proposition~\ref{prop_wiretap-set}, it follows that
\begin{align}\label{equ-add-1}
\mincut(s, A_2)=\mincut(s, A_1'\cup A_2),
\end{align}
i.e., a minimum cut between $s$ and $A_1'\cup A_2$ is actually a minimum cut between $s$ and $A_2$. Define $\MinCut(A_1'\cup A_2)$ as the set of all the minimum cuts between $s$ and $A_1'\cup A_2$. Then for any $A_1'\in \Cl(A_1)$, by (\ref{equ-add-1}) we have $A_2 \sim B$ for every minimum cut $B\in \MinCut(A_1'\cup A_2)$. With a slight abuse of notation, denote this by $\MinCut(A_1'\cup A_2)\sim A_2$.

Let $\CUT$ be a common minimum cut of the wiretap sets in $\Cl(A_2)$. It follows that $\CUT \sim A_2 \sim \MinCut(A_1'\cup A_2)$ for every $A_1'\in \Cl(A_1)$. Now, for each $A_1'\in \Cl(A_1)$, choose $\CUT_{A_1'} \in \MinCut(A_1'\cup A_2)$ arbitrarily. Then $\CUT \sim \CUT_{A_1'}$ for every $A_1'\in \Cl(A_1)$. By Proposition~\ref{lem_1}, we obtain
\begin{align*}
\mincut\bigg(s, \CUT \cup \bigcup_{A_1'\in\Cl(A_1)} \CUT_{A_1'}\bigg)=\mincut(s, \CUT)=|A_2|,
\end{align*}
which implies that there exists a common minimum cut $\CUT^*$ of cardinality equal to $|A_2|$ separating $\CUT$ and $\CUT_{A_1'}$ for every $A_1'\in\Cl(A_1)$. Therefore, $\CUT^*$ is a common minimum cut of the wiretap sets in $\Cl(A_2)$ which also separates each of the wiretap sets in $\Cl(A_1)$, i.e., $\Cl(A_1)\prec \Cl(A_2)$ by Definition~\ref{defn_cl-domin}. Theorem~\ref{thm_cl-domination} is proved.


\section{Proof of Lemma~\ref{lem_2}}\label{app:b}

Let $|A_1|=r_1$, $|A_2|=r_2$ and clearly $r_1<r_2$ since $A_1\prec A_2$. Since $A_1$ is regular and $\CUT_1\in \MinCut(A_1)$, it follows that
\begin{align}\label{equ_app_1}
|\CUT_1|=\mincut(s, A_1)=|A_1| = r_1.
\end{align}
Further since $A_1\prec A_2$ and $\CUT_{1,2}\in \MinCut(A_1\cup A_2)$, by Proposition~\ref{prop_wiretap-set} we have
\begin{align}\label{equ_app_1_1}
|\CUT_{1,2}|=\mincut(s, A_1\cup A_2)=\mincut(s, A_2)=|A_2|=r_2.
\end{align}

To prove $\CUT_1 \prec \CUT_{1,2}$, by Proposition~\ref{prop_wiretap-set}, it suffices to prove that
$$\mincut(s, \CUT_1\cup \CUT_{1,2})=\mincut(s, \CUT_{1,2})=r_2.$$
First note that
\begin{align}\label{equ_app_2}
\mincut(s, \CUT_1\cup \CUT_{1,2})\geq \mincut(s, A_1\cup A_2)=r_2,
\end{align}
where the inequality follows from the fact that a cut between $s$ and $\CUT_1\cup \CUT_{1,2}$ separates $A_1\cup A_2$ from $s$. Hence, we only need to prove that $\mincut(s, \CUT_1\cup \CUT_{1,2})\leq r_2$.

Let $\mincut(s, \CUT_1\cup \CUT_{1,2})=r$. Then there exist $r$ edge-disjoint paths from $s$ to the edges in $\CUT_1\cup \CUT_{1,2}$, say $P_1, P_2, \cdots, P_r$, such that each path passes through exactly one edge in $\CUT_1\cup \CUT_{1,2}$ as the last edge of the path. This is explained as follows. Since the $r$ last edges of the $r$ paths are included in $\CUT_1\cup \CUT_{1,2}$, if one path of them passes through more than one edge in $\CUT_1\cup \CUT_{1,2}$, we can replace the path by its subpath from $s$ to the first edge on the path in $\CUT_1\cup \CUT_{1,2}$, and this new path is still edge-disjoint with the other $r-1$ paths and it passes through exactly one edge in $\CUT_1\cup \CUT_{1,2}$, i.e., the last edge of the new path.

Let $a$ ($a\leq r_1$ by (\ref{equ_app_1})) be the number of paths among the $r$ edge-disjoint paths $P_1, P_2, \cdots, P_r$ such that their last edges are in $\CUT_1\setminus\CUT_{1,2}$. Then for the remaining $r-a$ paths, the $r-a$ last edges of them are in $\CUT_{1,2}$. Without loss of generality, assume the former $a$ paths be $P_1, P_2, \cdots, P_a$ with the last edges being $e_1, e_2, \cdots, e_a \in \CUT_1\setminus\CUT_{1,2}$, respectively, and the latter $r-a$ paths be $P_{a+1}, P_{a+2}, \cdots, P_r$ with the last edges being $e_{a+1}, e_{a+2}, \cdots, e_r \in \CUT_{1,2}$, respectively. Let
\begin{align*}
\In(\CUT_1)&=\{ e_1, e_2, \cdots, e_a \}\subseteq \CUT_1\setminus\CUT_{1,2},\\
\In(\CUT_{1,2})&=\{ e_{a+1}, e_{a+2}, \cdots, e_r \}\subseteq \CUT_{1,2},
\end{align*}
and
\begin{align*}
&\overline{\In(\CUT_{1,2})}=\CUT_{1,2} \setminus \In(\CUT_{1,2}).
\end{align*}
Then
\begin{align}\label{equ8}
|\In(\CUT_1)|+|\In(\CUT_{1,2})|=r.
\end{align}

Since $\CUT_1$ is a minimum cut between $s$ and the wiretap set $A_1$, i.e., $|\CUT_1|=\mincut(s, A_1)=|A_1|$, there are $r_1$ edge-disjoint paths from $\CUT_1$ to $A_1$ that start with all the $r_1$ distinct edges in $\CUT_1$ and end with all the $r_1$ distinct edges in $A_1$. Denote such $r_1$ paths by $P_1', P_2', \cdots, P_{r_1}'$ and without loss of generality assume that $P_1', P_2', \cdots, P_{a}'$ start with $e_1, e_2, \cdots, e_a$, respectively. Note that $P_i\cap P_i'=\{e_i\}$ for all $1\leq i \leq a$, since the network $G$ is acyclic. Next, we prove by contradiction that
\begin{align}\label{equ2}
P_i' \cap \CUT_{1,2} \neq \emptyset, \qquad \forall\ 1\leq i \leq a.
\end{align}
Assume $P_i' \cap \CUT_{1,2} = \emptyset$ for some $i$, $1\leq i \leq a$. Since the path $P_i$ from $s$ to $e_i$ does not contain any edge in $(\CUT_1\cup \CUT_{1,2})\setminus\{e_i\}$, $P_i\cup P_i'$ constitutes a path from $s$ to some edge in $A_1$ not including any edge in $\CUT_{1,2}$, which contradicts to the assumption that $\CUT_{1,2}$ separates $A_1$ from $s$. Hence, we have proved (\ref{equ2}).

We further prove by contradiction that
\begin{align}\label{equ3}
P_i' \cap \overline{\In(\CUT_{1,2})}\neq \emptyset, \qquad \forall\ 1\leq i \leq a.
\end{align}
Suppose $P_i' \cap \overline{\In(\CUT_{1,2})}=\emptyset$ for some $i$, $1\leq i \leq a$. Note that $P_i'$ does not pass through any edge in $\CUT_1\backslash \{e_i\}$. Together with (\ref{equ2}), $P_i'$ must pass through an edge in $\In(\CUT_{1,2})\setminus \CUT_1$. Consider the last edge in $\In(\CUT_{1,2})\setminus \CUT_1$ that $P_i'$ pass through. Without loss of generality, let this edge be $e_{a+1}$. Then
\begin{align}\label{equ-lem-6}
e_{a+1}\notin \CUT_1.
\end{align}
Thus, the subpath of $P_i'$ from $e_{a+1}$ to some edge in $A_1$ does not contain any edge in $(\CUT_1\cup \CUT_{1,2})\setminus\{ e_{a+1}\}$. Recall from the foregoing that the path $P_{a+1}$ does not contain any edge in $(\CUT_1\cup \CUT_{1,2})\setminus\{e_{a+1}\}$. Then together with (\ref{equ-lem-6}), we see that concatenating $P_{a+1}$ and the subpath of $P_i'$ from $e_{a+1}$ to some edge in $A_1$ yields a path from $s$ to some edge in $A_1$ without passing through any edge in $\CUT_1$. This contradicts the assumption that $\CUT_1$ is a minimum cut between $s$ and $A_1$. Hence, we have proved~(\ref{equ3}).

Now, $P_i'$, $1\leq i \leq a$, are edge-disjoint. Together with (\ref{equ3}), we have $|\overline{\In(\CUT_{1,2})}|\geq a$, or equivalently, $|\In(\CUT_{1,2})|\leq r_2-a$. It then follows from $|\In(\CUT_1)|=a$ and (\ref{equ8}) that
\begin{align*}
r=|\In(\CUT_1)|+|\In(\CUT_{1,2})|\leq a+r_2-a=r_2,
\end{align*}
that is, $\mincut(s, \CUT_1\cup \CUT_{1,2})\leq r_2$. Lemma~\ref{lem_2} is proved.

\section{Proofs of Lemma~\ref{prop1_relation-cuts} and Theorem~\ref{thm_relation-cuts}}\label{app:c}

\begin{IEEEproof}[Proof of Lemma~\ref{prop1_relation-cuts}]
The ``only if'' part of the lemma is trivial.

We now prove the ``if'' part. Let $\CUT_1=\{e_{1,i}:\ i=1, \ldots, n  \}$ and $\CUT_2=\{e_{2,i}:\ i=1, \ldots, n  \}$ be two minimum cuts in $\MinCut(t)$. Let $P_1, P_2, \cdots, P_n$ be $n$ edge-disjoint paths from $s$ to $t$ such that for each $i$, $1 \leq i \leq n$, $P_i \cap \CUT_1=\{ e_{1,i} \}$, $P_i \cap \CUT_2=\{ e_{2,i} \}$, and $e_{1,i}\leq e_{2,i}$. We now prove the ``if'' part by contradiction. Assume the contrary that $\CUT_1 \leq \CUT_2$ is false, i.e., $\CUT_1$ is not a cut separating $\CUT_2$ from $s$. Upon deleting the edges in $\CUT_1$, there still exists a path, say $P$, from $s$ to an edge in $\CUT_2$, say $e_{2,1}$ ($P$ includes $e_{2,1}$). Note that the path $P$ and the subpath of $P_1$ from $\head(e_{2,1})$ to $t$ are edge-disjoint since the network is acyclic. In addition, the subpath of $P_1$ from $\head(e_{2,1})$ to $t$ does not contain the edges in $\CUT_1$ since $P_1 \cap \CUT_1=\{e_{1,1}\}$ and $e_{1,1}\leq e_{2,1}$. Hence, concatenating $P$ and the subpath of $P_1$ from $\head(e_{2,1})$ to $t$ yields a new path from $s$ to $t$ that contains no edges in $\CUT_1$, which contradicts the assumption that $\CUT_1\in \MinCut(t)$. The proof is completed.
\end{IEEEproof}

\begin{IEEEproof}[Proof of Theorem~\ref{thm_relation-cuts}]
We will prove the theorem by contradiction. Suppose that there exist $n$ edge-disjoint paths from $s$ to $t$, denoted by $P_1$, $P_2$, $\cdots$, $P_n$, such that one of them, say $P_1$, passes through an edge $e_{1,1} \in \CUT_1$ and an edge $e_{2,1} \in \CUT_2$ (i.e., $P_1 \cap \CUT_1=\{e_{1,1}\}$ and $P_1 \cap \CUT_2=\{e_{2,1}\}$) with $e_{2,1} < e_{1,1}$ (i.e., $e_{2,1} \leq e_{1,1}$ and $e_{2,1} \neq e_{1,1}$). Now, we divide the path $P_1$ into two disjoint subpaths: the subpath from $s$ to $e_{2,1}$ (including $e_{2,1}$), and the subpath from $\head(e_{2,1})$ to $t$ passing through $e_{1,1}$. By Proposition~\ref{prop}, the first subpath of $P_1$ from $s$ to $e_{2,1}$ contains no edges in $(\CUT_1\cup \CUT_2)\setminus \{e_{2,1}\}$. In other words, there exists a path from $s$ to an edge in $\CUT_2$ (i.e., $e_{2,1}$) upon deleting all the edges in $\CUT_1$, a contradiction to $\CUT_1 \leq \CUT_2$. The theorem is proved.
\end{IEEEproof}

\section{Verification of Algorithm~\ref{algo_Augmenting-Path}}\label{app:d}

In this appendix, we verify that the output edge set $\CUT$ of Algorithm~\ref{algo_Augmenting-Path} is the primary minimum cut between $s$ and $t$. We adopt the standard terminologies in network flow theory. In a network $G$ with a flow $f$, a non-source node $u$ is called {\em reachable} from $s$ if there exists an {\em $f$-unsaturated path} from $s$ to $u$, where an $f$-unsaturated path means that each edge $e$ on this path is either a forward edge with flow value $0$ or a reverse edge with flow value $1$. For a detailed discussion on unsaturated path, we refer the reader to \cite[Chapter 7]{Book-GraphTh-Bondy-Murty}. The following lemma is also standard.

\begin{lemma}\label{lem:unsaturated-path}\em
In a network $G$ with a flow $f$ from the source node $s$ to a sink node $t$, if there exists an $f$-unsaturated path from $s$ to $t$, then by ``flipping'' this path, i.e., replacing the flow value $0$ of the forward edges on the path by $1$ and the flow value $1$ of the reverse edges on the path by $0$, a new flow $f'$ is obtained and the flow value of $f'$ is increased by $1$, i.e., $v(f')=v(f)+1$. In particular, if no unsaturated paths from $s$ to $t$ exist, the flow is a maximum flow from $s$ to $t$.
\end{lemma}

Let $\CUT=\{e_i:\ 1\leq i \leq n\}$ be the output edge set of Algorithm~\ref{algo_Augmenting-Path}. Then the nodes $\tail(e_i)$, $1\leq i \leq n$ are reachable and the nodes $\head(e_i)$, $1\leq i \leq n$ are unreachable from $s$. This further implies that all the edges in $\CUT$ have flow value $1$, i.e., $f(e_i)=1$, $1\leq i \leq n$, because otherwise $\head(e_i)$ would be included in the set $S$ when the algorithm terminates.

First, we can easily see that $\CUT$ is indeed a cut between $s$ and $t$, i.e., $t\notin S$, because otherwise there exists an unsaturated path from $s$ and $t$, implying that $f$ is not a maximum flow by Lemma~\ref{lem:unsaturated-path}. It follows that $n\geq C_t \geq 1$.

We now prove that $\CUT$ is minimum, i.e., $n=C_t$. Assume the contrary that $n>C_t$. Then the maximum flow $f$ can be decomposed into $C_t$ edge-disjoint paths $P_1, P_2, \cdots, P_{C_t}$ from $s$ to $t$ with
\begin{align}\label{equ:flow_value}
f(e)=\left\{
  \begin{array}{ll}
    1, & \hbox{$e\in P_i$ for some $1\leq i \leq C_t$;} \\
    0, & \hbox{otherwise.}
  \end{array}
\right.
\end{align}
Since $f(e_i)=1$ for $1\leq i \leq n$, each $e_i$ must be on one of the $C_t$ edge-disjoint paths from $s$ to $t$. Furthermore, since $n>C_t$, there exists a path $P_j$ that contains at least $2$ edges in $\CUT$, say $e_1$ and $e_2$. We assume without loss of generality that $e_1\leq e_2$ on $P_j$. Note that $\tail(e_2)$ is reachable from $s$. If $\tail(e_2)=\head(e_1)$, then $\head(e_1)$ is reachable from $s$, which is a contradiction because $e_1\in \CUT$. Otherwise, let $\hat{e}$ be the predecessor of $e_2$ on $P_j$. Since $f(\hat{e})=1$, $\tail(\hat{e})$ is also reachable from $s$ (through $\tail(e_2)$). By repeating this argument if necessary, we see inductively that $\head(e_1)$ is reachable from $s$, a contradiction. Therefore, $\CUT$ must be a minimum cut between $s$ and $t$, i.e., $n=C_t=v(f)$.

It remains to prove that $\CUT$ is primary. Assume that $\CUT$ is not primary, and instead let $\CUT^*=\{e_i^*:\ 1\leq i \leq n\}$ be the primary minimum cut between $s$ and $t$. By Definitions~\ref{defn_primary-mincut} and \ref{defn_relation-cuts}, we have $\CUT^* \leq \CUT$. By Proposition~\ref{prop}, we can let $\CUT\cap P_i=\{e_i\}$ and $\CUT^* \cap P_i=\{e_i^*\}$ for $1\leq i \leq n$. Then $e_i^*\leq e_i$ for all $1\leq i \leq n$ by Theorem~\ref{thm_relation-cuts}, which implies that for each $1\leq i \leq n$, the subpath of $P_i$ from $\head(e_i)$ to $t$ contains no edges in $\CUT^*$. Since we assume that $\CUT\neq \CUT^*$, there exists $1\leq i \leq n$ such that $e_i\neq e_i^*$. Without loss of generality assume that $e_1\neq e_1^*$, and let $P$ be an  $f$-unsaturated path from $s$ to $\tail(e_1)$.

Now, consider any edge $e\in P\cap P_1$. Since $e\in P_1$, we have $f(e)=1$, which together with $e\in P$ implies that $e$ must be a reverse edge on $P$. Thus, we have proved the following claim which will be used throughout the rest of the proof.
\begin{claim}\label{claim:1}\em
For every edge $e\in P\cap P_1$, $\head(e)$ is the node on $P$ immediately before $\tail(e)$.
\end{claim}

We now prove by contradiction that $P$ and the subpath of $P_1$ from $\tail(e_1)$ to $t$, denoted by $P_1^{\tail(e_1)\rightarrow t}$,\footnote{Let $P$ be an (unsaturated) path from $s$ to a non-source node $u$. For any two nodes $u_1$ preceding $u_2$ on $P$, the subpath of $P$ from $u_1$ to $u_2$ is denoted by $P^{u_1\rightarrow u_2}$ throughout this proof to simplify notation.} are edge-disjoint. Let $e\in P \cap P_1^{\tail(e_1)\rightarrow t}$. We consider two cases:
\newline\underline{$e=e_1$} By Claim~\ref{claim:1}, we can see that $\tail(e_1)$ is reachable from $s$ on $P$ through $\head(e_1)$, which implies that $\head(e_1)$ is also reachable from $s$, a contradiction.
\newline\underline{$e\neq e_1$} Since $e\in P\cap P_1^{\head(e_1)\rightarrow t}$, $\tail(e)$ is reachable from $s$ (through $\head(e)$) since $e$ is on $P$. Together with the flow value of each edge (if exists) on the subpath $P_1^{\head(e_1)\rightarrow \tail(e)}$ being $1$, by the argument previously used in proving that $\CUT$ is minimum, $\head(e_1)$ is also reachable from $s$, which again is a contradiction.

We now prove by contradiction that $\CUT^*$ is not the primary minimum cut between $s$ and $t$ by considering two cases.

\noindent\textbf{Case 1:} $P\cap P_1^{s \rightarrow \tail(e_1)}=\emptyset$.

We will prove that in this case a new maximum flow $f''$ with $f''(e_1^*)=0$ can be found, i.e., the $n$ edge-disjoint paths from $s$ to $t$ with respect to $f''$ does not pass through $e_1^*$. First, we define $f'$ as
$$f'(e)=\left\{
  \begin{array}{ll}
    0, & \hbox{$e\in P_1$;} \\
    f(e), & \hbox{otherwise;}
  \end{array}
\right.$$
which is a flow but no longer a maximum flow since $v(f')=v(f)-1=n-1$.
Then we can obtain an $f'$-unsaturated path $\hat{P}_1$ from $s$ to $t$ by concatenating $P$ and $P_1^{\tail(e_1) \rightarrow t}$. Since $e_1^*<e_1$ on $P_1$, $e_1^*$ is on the subpath $P_1^{s\rightarrow \tail(e_1)}$. Together with $P\cap P_1^{s \rightarrow \tail(e_1)}=\emptyset$, we have $e_1^*\notin \hat{P}_1$. Now, define a flow $f''$ by flipping the flow values in the $f'$-unsaturated path $\hat{P}_1$, i.e.,
\begin{align}\label{equ:flip-path}
f''(e)=\left\{
  \begin{array}{lll}
    1, & \hbox{$e\in \hat{P}_1$ with $f'(e)=0$;} \\
    0, & \hbox{$e\in \hat{P}_1$ with $f'(e)=1$;} \\
 f'(e), & \hbox{otherwise.}
  \end{array}
\right.
\end{align}
By Lemma~\ref{lem:unsaturated-path}, we have $v(f'')=v(f')+1=n$.
We then have obtained a maximum flow $f''$ with $f''(e_1^*)=0$. By Proposition~\ref{prop}, $\CUT^*$ is not a minimum cut between $s$ and $t$, and hence not the (primary) minimum cut between $s$ and $t$.

\noindent\textbf{Case 2:} $P\cap P_1^{s \rightarrow \tail(e_1)}\neq\emptyset$.

Let $e'$ be the first edge on $P$ that is also on $P_1$. We consider two cases.

\textbf{Case 2A:} $e_1^*\leq e'$.

By Claim~\ref{claim:1}, $e'$ is a reverse edge on $P$ with $f(e')=1$ and $e'<e_1$. Then the $f$-unsaturated subpath $P^{s \rightarrow \head(e')}$ does not pass through $\tail(e')$, and hence does not contain $e'$. On the other hand, since $e'$ is the first edge on $P$ that is also on $P_1$, $P^{s \rightarrow \head(e')}$ does not contain any edge on $P_1$. Thus, $P^{s \rightarrow \head(e')}$ is edge-disjoint with $P_1$, and therefore also with the subpath $P_1^{s \rightarrow \head(e')}$. Since $e_1^*\leq e'$, $e_1^*$ is on $P_1^{s \rightarrow \head(e')}$ and hence not on $P^{s \rightarrow \head(e')}$ and $P_1^{\head(e') \rightarrow t}$. By considering the concatenation of $P^{s \rightarrow \head(e')}$ and $P_1^{\head(e') \rightarrow t}$, we see by using the same argument as in Case 1 that $\CUT^*$ is not the primary minimum cut between $s$ and $t$.

\textbf{Case 2B:} $e_1^*>e'$.

Let $\tilde{e}$ be the last edge on $P\cap P_1$ such that $\tilde{e}<e_1^*$. Consider the following two cases:
      \begin{enumerate}
        \item $P^{\tail(\tilde{e}) \rightarrow \tail(e_1)} \cap P_1^{\tail(\tilde{e}) \rightarrow \tail(e_1)}=\emptyset$. Since $\tilde{e}<e_1^*<e_1$, $e_1^*\in P_1^{\tail(\tilde{e}) \rightarrow \tail(e_1)}$ and hence $P^{\tail(\tilde{e}) \rightarrow \tail(e_1)}$ does not contain $e_1^*$. On the other hand, since $\tilde{e}$ is the last edge on $P\cap P_1$ such that $\tilde{e}<e_1^*$, $P^{\tail(\tilde{e}) \rightarrow \tail(e_1)}$ contains no edges on $P_1$, where we note that $\tilde{e}$ is not on $P^{\tail(\tilde{e}) \rightarrow \tail(e_1)}$ by Claim~\ref{claim:1}. Then $P^{\tail(\tilde{e}) \rightarrow \tail(e_1)}$ is edge-disjoint with $P_1$. By considering the concatenation of $P_1^{s \rightarrow \tail(\tilde{e})}$, $P^{\tail(\tilde{e}) \rightarrow \tail(e_1)}$, and $P_1^{\tail(e_1) \rightarrow t}$, we see by using the same argument as in Case 1 that $\CUT^*$ is not the primary minimum cut between $s$ and $t$.
        \item $P^{\tail(\tilde{e}) \rightarrow \tail(e_1)} \cap P_1^{\tail(\tilde{e}) \rightarrow \tail(e_1)}\neq \emptyset$. Let $\hat{e}$ be the first edge on $P \cap P_1$ such that $\hat{e}\geq e_1^*$. Together with $\tilde{e}$ being the last edge on $P\cap P_1$ such that $\tilde{e}<e_1^*$, the subpath $P^{\tail(\tilde{e}) \rightarrow \head(\hat{e})}$ contains no edges on $P_1$ by Claim~\ref{claim:1}, and hence $P^{\tail(\tilde{e}) \rightarrow \head(\hat{e})}$ is edge-disjoint with $P_1$. On the other hand, we note that $\tilde{e}<e_1^*\leq\hat{e}<e_1$ on $P_1$, implying that $e_1^*\in P_1^{\tail(\tilde{e}) \rightarrow \head(\hat{e})}$. Thus, considering the concatenation of $P_1^{s \rightarrow \tail(\tilde{e})}$, $P^{\tail(\tilde{e}) \rightarrow \head(\hat{e})}$, and $P_1^{\head(\hat{e}) \rightarrow t}$, we see by using the same argument as in Case 1 that $\CUT^*$ is not the primary minimum cut between $s$ and $t$.
\end{enumerate}

Combining all the above, Algorithm~\ref{algo_Augmenting-Path} is verified.


\end{document}